\documentclass[rsi,amsmath,amssymb,reprint,pre]{revtex4-1}
\usepackage[T1]{fontenc}
\usepackage[english]{babel}
\usepackage{graphicx}
\usepackage{dcolumn}
\usepackage{bm}

\DeclareMathAlphabet{\mathbfit}{OML}{cmm}{b}{it}

\begin{document}

\preprint{AIP/123-QED}

\title{Lattice-Boltzmann simulations of electrowetting phenomena}

\author{\'Elfego Ruiz-Guti\'errez}
\email{elfego.r.gutierrez@northumbria.ac.uk}
\affiliation{Smart Materials and Surfaces Laboratory, Northumbria University, Ellison Place, Newcastle upon Tyne, NE1 8ST, UK.}

\author{Rodrigo Ledesma-Aguilar}
\affiliation{Smart Materials and Surfaces Laboratory, Northumbria University, Ellison Place, Newcastle upon Tyne, NE1 8ST, UK.}

\date{\today}

\begin{abstract}
  We present a lattice-Boltzmann method that can simulate the coupled hydrodynamics and electrostatics equations of motion of a two-phase fluid as a means to model electrowetting phenomena.
  Our method has the advantage of modelling the electrostatic fields within the lattice-Boltzmann algorithm itself, eliminating the need for a hybrid method. 
  We validate our method by reproducing the static equilibrium configuration of a droplet subject to an applied voltage and show that the apparent contact angle of the drop depends on the voltage following the Young-Lippmann equation up to contact angles of $\approx 50^\circ$.  
  At higher voltages, we observe a saturation of the contact angle caused by the competition between electric and capillary stresses. 
  We also study the stability of a dielectric film trapped between a conducting fluid and a solid electrode and find a good agreement with analytical predictions based on lubrication theory.   % 
  Finally, we investigate the film dynamics at long times and report observations of film breakup and entrapment similar to previously reported experimental results. 
\end{abstract}

% \keywords{Lattice-Boltzmann simulations, Electrowetting}
\maketitle
%%% END OF TITLE, AUTHORS AND ABSTRACT%%%

%%%MAIN TEXT%%%%

\section{Introduction}
\label{sec:introduction}

Electrowetting refers to the spreading of an electrically conducting liquid on a solid electrode when a voltage difference is applied between the two~\cite{mugele2005electrowetting}. 
Because of its ability to control the interaction of liquids with solid surfaces, electrowetting has triggered a number of applications, such as droplet-based microfluidic devices~\cite{teh2008droplet,ren2002droplettransport}, droplet actuation~\cite{baret2005electroactuation} and mixing~\cite{mugele2006electromixing,lu2017dynamics,kudina2016emaldi}, deformable optical apertures~\cite{schuhladen2016optofluidic} and lenses~\cite{berge2000electrolensing}, and electronic paper displays~\cite{hayes2003epaperspeed,you2010epaper3color}.
Broadly speaking, there are two types of electrowetting setups: Electrowetting On Conductor (EWOC), in which the conductive liquid is in direct contact with the solid electrode~\cite{lomax2016ewoc}, and the more popular Electrowetting On Dielectric (EWOD), in which direct contact is removed by coating the electrode with a dielectric layer~\cite{nelson2012droplet}.

The simplest electrowetting situation, used widely in many EWOC and EWOD setups, is the spreading of a droplet of conductive liquid suspended in an ambient dielectric fluid that completely wets the solid surface~\cite{quilliet2002ewinterfacepot}.
During the actuation, the ambient fluid forms a thin film underneath the droplet that can become unstable and break up into small ``bubbles'' that remain in contact with the solid~\cite{kuo2003ewmovement,staicu2006entrapment}.
Such a transition introduces mobile contact lines~\cite{mugele2009fundamental}, which can drastically affect the friction force acting on its overall dynamics~\cite{mchale2013spreading}.
On the other hand, the spreading of a droplet at high voltages can reach a saturation regime, where the apparent contact angle that the droplet forms with the solid settles to a limiting value~\cite{buehrle2003profiles}. 
At even higher voltages, the edge of the spreading droplet can become unstable, and trigger the breakup of small droplets that form coronal patterns around the mother drop \cite{quilliet2001cas}.

Despite these important advances, the rich phenomenology of electrowetting remains to be fully understood.  
For this purpose, it is essential to develop computational methods that capture the multiphase fluid dynamics and that resolve the effect of electrostatic interactions, as these can help interpret experiments and inform theory. 
The Lattice-Boltzmann Method (LBM) has proved to be a powerful tool to study mulitphase fluid dynamics~\cite{kruger2016lattice}.
To implement electrowetting within the LBM, it has been proposed to prescribe the interaction energy of the surface~\cite{li2009lbmew,clime2010numericalew}, which leads to an effective   contact angle.  
Such an approach, however, does not capture the underlying coupling between the hydrodynamic and electrostatic fields. 
As a means to overcome this limitation, hybrid methods that solve the electrostatic field equations separately have been developed~\cite{aminfar2009lbmew}, but these come at the expense of running and coupling two numerical solvers concurrently. 

Here we present a lattice-Boltzmann method capable of solving the coupled hydrodynamics-electrostatics equations that govern electrowetting phenomena within a single algorithm. 
We use the so-called free-energy approach as a starting point to model the multiphase fluid dynamics, and show that the effect of the electrostatic energy can be included explicitly in the corresponding energy functional. 
We introduce a set of lattice-Boltzmann equations, where the electrostatic potential field is determined by a new set of distribution functions. 
We validate this ``all-in-one'' method by comparing the electrowetting-induced spreading of a droplet to the classical theory of Young and Lippman~\cite{lippmann1875relations}. 
To illustrate the utility of the method, we present results of the stability of the thin film separating a conducting droplet and a solid electrode, considering both the linear and non-linear regimes.

% end of section introduction

\section{Diffuse-interface model of electrowetting phenomena} % (fold)
\label{sec:diffuse_interface_model}

Let us consider two incompressible,  immiscible fluids: a perfect conductor, corresponding to the spreading liquid,
and a dielectric, corresponding to the surrounding phase.
We describe the two-fluid system using a diffuse-interface model that identifies each phase using an order parameter, or phase field, $\phi ({\mathbfit{x}},t)$, where $\mathbfit{x}$ denotes the position vector and $t$ denotes time. 
Without loss of generality, we let $\phi > 0$ be the conductive phase and $\phi < 0 $ be the dielectric.

The Helmholtz free energy of the fluid-fluid system can be defined as~\cite{landau1980statistical}
\begin{equation}
		F_{\rm th}[\phi] := \int_\Omega \psi(\phi, \mathbfit{\nabla} \phi) \,\mathrm{d}^3x + \int_{\partial \Omega} \zeta \, \phi \,\mathrm{d} S.
	\label{eq:helmholtz_free_energy}
\end{equation}
The first term corresponds to the volumetric contribution to the free energy over the region occupied by the fluid, $\Omega$.  
This consists of the well-known energy density of a binary fluid~\cite{bray1994theory,cahn1958free-i},
\begin{equation}
		\psi(\phi, \mathbfit{\nabla} \phi) := \frac{3 \gamma}{\sqrt{8} \ell} \left[ \frac{\phi^4}{4} - \frac{\phi^2}{2} + \frac{\ell^2}{2} | \mathbfit{\nabla} \phi |^2 \right] ,
	\label{eq:volume_free_energy}
\end{equation}
where the square-gradient term allows the coexistence of the two bulk phases, of equilibrium phase-field values $\phi=\pm1$, separated by a diffuse interface of thickness $\ell$ and surface tension $\gamma$. 
The second integral in Eq.~(\ref{eq:helmholtz_free_energy}) corresponds to the surface interaction energy of the fluid with the solid electrode, whose boundary is denoted by $\partial \Omega$,
and where the constant $\zeta$ is called the wetting potential~\cite{cahn1977critical}. 

In equilibrium, and in the absence of an electric field, the fluid-fluid interface is expected to intersect the solid boundary at an angle $\theta_0$ determined by the Young-Dupr\'e relation~\cite{cahn1977critical},
\begin{equation}
	\gamma_\text{sd} - \gamma_\text{sc} = \gamma \cos \theta_0,
	\label{eq:young-dupre}
\end{equation}
where $\gamma_\text{sd}$ and $\gamma_\text{sc}$ are the solid-dielectric fluid and solid-conductive fluid surface tensions.
This is a standard result that can be obtained from Eqs.~(\ref{eq:helmholtz_free_energy})--(\ref{eq:young-dupre}), which yield a relation between the wetting potential and the contact angle~\cite{briant2004contact-i}:
\begin{equation}
  \zeta = \frac{3}{2} \gamma \mathrm{sgn}(\theta_0 - \pi / 2) \left\{ \cos(\alpha / 3) \left[ 1 - \cos(\alpha / 3) \right] \right\}^{1/2},
  \label{eq:wetting_potential}
\end{equation}
where $\alpha = \arccos ( \sin^2 \theta_0 )$.  
It can also be shown that, in such a limit, the pressure field, $p(\mathbfit{x})$, is uniform in each phase, but jumps across the interface satisfying the Young-Laplace relation  
\begin{equation}
	\Delta p = 2 \gamma \kappa,
	\label{eq:young-laplace}
\end{equation}
where $\kappa$ is the interface curvature~\cite{yang1976molecular}.

To model the electrostatic behaviour of the fluid mixture we introduce the electrostatic free energy:
\begin{equation}
		F_{\rm el}[V] := - \frac{1}{2} \varepsilon \int_\Omega  |\mathbfit{E} |^2 \,\mathrm{d}^3x,
	\label{eq:helmholtz_free_energy_el}
\end{equation}
which quantifies the potential energy density of the electric field $\mathbfit{E}({\mathbfit{x}}) = - \mathbfit{\nabla} V$, where $V({\mathbfit{x}})$ is the electric potential and $\varepsilon$ is the electric permittivity~\cite{landau2013electrodynamics,jackson1999classical}.

Out of equilibrium, local differences in the total free energy, $F = F_\text{th} + F_\text{el}$, give rise to capillary and electrostatic forces. 
On the one hand, changes in the phase field lead to a chemical potential field
\begin{equation}
	\vartheta ({\mathbfit{x}},t) := \frac{\delta F}{\delta \phi} = \frac{3 \gamma}{\sqrt{8} \ell} \left[ \phi (\phi^2 - 1) - \ell^2 \nabla^2 \phi \right],
	\label{eq:chemical_potential}
\end{equation}
and a corresponding capillary force density 
\begin{equation}
	\mathbfit{f}_\text{cap} = - \phi \mathbfit{\nabla} \vartheta,
	\label{eq:chemical_force}
\end{equation}
which reduces to Eq.~(\ref{eq:young-laplace}) in equilibrium~\cite{yang1976molecular}.
On the other hand, changes in the electric potential give rise to the electric charge distribution~\cite{jackson1999classical}
\begin{equation}
	\varrho_\text{el} ({\mathbfit{x}},t) := - \frac{\delta F}{\delta V} = - \varepsilon \nabla^2 V = \varepsilon \mathbfit{\nabla} \cdot \mathbfit{E},
	\label{eq:electric_charge_density}
\end{equation}
and to the electric force density
\begin{equation}
	\mathbfit{f}_\text{el} = \varrho_\text{el} \mathbfit{E},
	\label{eq:electrical_force}
\end{equation}
which is the Lorentz force in the absence of magnetic fields~\cite{jackson1999classical}.

The chemical and electrostatic force densities, Eqs.~(\ref{eq:chemical_force}) and~(\ref{eq:electrical_force}), together with the local pressure gradient, $-\nabla p$, change the momentum of the fluid.
The resulting total force density can be written in terms of a generalised pressure tensor, $\boldsymbol{\Pi}$, i.e., 
\begin{equation}
	-\mathbfit{\nabla} \cdot \boldsymbol{\Pi} := -\nabla p + \mathbfit{f}_\text{cap} + \mathbfit{f}_\text{el}.
	\label{eq:pressure_just}
\end{equation}
This leads to the expression
\begin{equation}
	\boldsymbol{\Pi} = \left(\phi \, \vartheta - \psi\right) \mathbf{I} + \frac{3 \gamma \ell}{\sqrt{8}} \mathbfit{\nabla} \phi \mathbfit{\nabla} \phi - \varepsilon \left( \mathbfit{E} \mathbfit{E} - \frac{1}{2} |\mathbfit{E}|^2 \mathbf{I} \right),
  \label{eq:generalised_pressure_tensor}
\end{equation}
where the last term in brackets is the Maxwell stress tensor~\cite{jackson1999classical} and $\mathbf{I}$ is the identity matrix.

The equations of motion of the fluids are obtained as follows. First, imposing the conservation of momentum leads to the incompressible Navier-Stokes equations
\begin{equation}
	\rho \left( \partial_t + \mathbfit{u} \cdot \mathbfit{\nabla} \right) \mathbfit{u} = - \nabla \cdot \left[ \boldsymbol{\Pi} - \frac{\mu}{2} \left( \nabla \mathbfit{u} + \nabla \mathbfit{u}^\text{T} \right) \right],
	\label{eq:Navier-Stokes}
\end{equation}
where $\mathbfit{u} (\mathbfit{x},t)$, $\rho$ and $\mu(\mathbfit{x})$ are the velocity field, density and dynamic viscosity of the fluid, respectively, and the superscript $\text{T}$ denotes matrix transposition.
To allow viscosity differences between the two phases we impose the local viscosity as
\begin{equation}
  \mu(\phi) := \frac{1 + \phi}{2} \mu_\text{c} + \frac{1- \phi}{2} \mu_\text{d},
  \label{eq:viscosity}
\end{equation}
where $\mu_\text{c}$ and $\mu_\text{d}$ are the bulk viscosities of the conductive and dielectric fluids.

Imposing the conservation of the phase field leads to a convection-diffusion equation, often referred to as the Cahn-Hilliard equation~\cite{swift1996lattice}:
\begin{equation}
	\partial_t \phi + \mathbfit{u} \cdot \mathbfit{\nabla}  \phi = M \nabla^2 \vartheta,
	\label{eq:Cahn-Hilliard}
\end{equation}
where $M$ is called the mobility.

To complete the formulation of the problem, we need to specify the electrostatic force density, which is a function of the potential field, $V$.
In the following, we assume that both phases are ideal, i.e., the conductor has a vanishing electrical resistivity, while the dielectric has a vanishing electrical conductivity.
It then follows that, since the electric field in the conductor is zero, the potential is constant in the bulk of that phase, i.e., 
\begin{equation}
  V = V_0 \quad \text{as }\phi\rightarrow 1.
  \label{eq:potential_conductor}
\end{equation}
On the other hand, for a perfect dielectric $\varrho_{\rm el}=0$, so Eq.~(\ref{eq:electric_charge_density}) reduces to
\begin{equation}
  \nabla^2 V = 0 \quad \text{as }\phi\rightarrow -1.
  \label{eq:harmonic_potential}
\end{equation}

The boundary conditions for the coupled set of PDEs, equations~(\ref{eq:Navier-Stokes}), (\ref{eq:Cahn-Hilliard}) and~(\ref{eq:harmonic_potential}), are specified as follows. 
For the velocity field we impose the impenetrability and no-slip boundary conditions:
\begin{equation}
\mathbfit{u}(\mathbfit{x}_\text{b}) = 0 \qquad \text{for~}\mathbfit{x}_\text{b}\in\partial\Omega.
\label{eq:No_Slip}
\end{equation}
For the phase field, we impose the natural boundary condition 
\begin{equation}
  \hat{\mathbfit{n}} \cdot \mathbfit{\nabla} \phi (\mathbfit{x}_\text{b}) = - \frac{\sqrt{8}}{3 \gamma \ell}\zeta(\theta_0) \qquad \text{for~}\mathbfit{x}_\text{b}\in\partial\Omega,
  \label{eq:Neumann_boundary_condition}
\end{equation}
where $\hat{\mathbfit{n}}$ is the unit normal to the solid boundary, and which enforces the wetting behaviour of the fluid-fluid mixture. 
Finally, for the potential we impose 
\begin{equation}
 V(\mathbfit{x}_\text{b}) = V_\text{b} \qquad \text{for~}\mathbfit{x}_\text{b}\in\partial\Omega,
  \label{eq:boundary_potential}
\end{equation}
where $V_\text{b}$ is the potential at the boundary.

% subsection diffuse_interface_model (end)

\section{Lattice-Boltzmann method} % (fold)
\label{sec:lattice_boltzmann_simulations}
In this section we formulate a lattice-Boltzmann algorithm capable of integrating Eqs.~(\ref{eq:Navier-Stokes}),~(\ref{eq:Cahn-Hilliard}) and~(\ref{eq:harmonic_potential}), 
subject to the boundary conditions~(\ref{eq:No_Slip})-(\ref{eq:boundary_potential}).

The lattice-Boltzmann method is a computational fluid dynamics solver that iterates the discretised Boltzmann equations
\begin{equation}
	f_q(\mathbfit{x} + \mathbfit{c}_q, t + 1) = f_q(\mathbfit{x}, t) + \mathcal{C}[f]_q
	\label{eq:lattice-Boltzmann_equation}
\end{equation}
and 
\begin{equation}
	g_q(\mathbfit{x} + \mathbfit{c}_q, t + 1) = g_q(\mathbfit{x}, t) + \mathcal{C}[g]_q, 
	\label{eq:lattice-Boltzmann}
\end{equation}
where $f_q$ and $g_q$ are particle distribution functions that represent the average number of fluid particles with position $\mathbfit{x}$ and velocity $\mathbfit{c}_q$ at time $t$.
Space and time are discretised, and the velocity space is sampled by a finite set of vectors $\{ \mathbfit{c} \}_{q=0}^{Q-1}$, where $Q$ is the number of directions in which the particle populations can move.
Here, we use the D2Q9 model, which consists of a two-dimensional square lattice with $Q=9$ (see Appendix~\ref{app:D2Q9}).

The time evolution of the distribution functions, given by Eqs.~(\ref{eq:lattice-Boltzmann_equation}) and~(\ref{eq:lattice-Boltzmann}), consists of a collision step and a streaming step.
The collision step, performed by the second term on the right-hand-side in each equation, relaxes the distribution functions local equilibrium values, $f_q^\text{eq}$ and $g_q^\text{eq}$.
Here we use the Multi-Relaxation Time scheme (MRT) to model the collision of the $f_q$, i.e.,
\begin{equation}
	\mathcal{C}[f]_q := - \sum_{r=0}^{Q-1} \Lambda_{q r} [f_r - f_r^\text{eq}](\mathbfit{x}, t),
	\label{eq:MRT}
\end{equation}
where the coefficients $\Lambda_{q r}$ determine the relaxation rate to equilibrium and are constructed using the Gram-Schmidt orthogonalisation procedure~\cite{dHumieres2002mrt}.
For the collision of the $g_q$ we use the single-relaxation time approximation, 
\begin{equation}
	\mathcal{C}[g]_q := -   \Lambda [g_q - g_q^\text{eq}](\mathbfit{x}, t),
	\label{eq:MRTg}
\end{equation}
where we set $\Lambda=1$, which helps improve the stability of the numerical method without loss of generality~\cite{kruger2016lattice}.

The connection between the lattice-Boltzmann equations and the hydrodynamic equations is done by relating the moments of the distribution functions to the hydrodynamic fields. 
The local mass, momentum and phase fields correspond to
\begin{equation}
	\rho = \sum_{q=0}^{Q-1} f_q,
	\label{eq:moment-0}
\end{equation}
\begin{equation}
	\rho \mathbfit{u} = \sum_{q=0}^{Q-1} \mathbfit{c}_q f_q
	\label{eq:moment-1}
\end{equation}
and 
\begin{equation}
	\phi = \sum_{q=0}^{Q-1} g_q. 
	\label{eq:phase_m0}
\end{equation}

The equilibrium distributions, $f_q^\text{eq}$ and $g_q^\text{eq}$, are constructed to convey the thermodynamic behaviour of the fluid and to ensure the local conservation of mass and momentum.
This is done by requiring that their moments satisfy the conditions: 
$\sum_q f_q^\text{eq} = \rho$, 
$\sum_q g_q^\text{eq} = \phi$,
$\sum_q \mathbfit{c}_q f_q^\text{eq} = \rho \mathbfit{u}$, 
$\sum_q \mathbfit{c}_q g_q^\text{eq} = \phi \mathbfit{u}$,
$\sum_q \mathbfit{c}_q \mathbfit{c}_q f_q^\text{eq} =  \boldsymbol{\Pi} + \rho \mathbfit{u} \mathbfit{u}$ and
$\sum_q \mathbfit{c}_q \mathbfit{c}_q g_q^\text{eq} = 2 M \vartheta \mathbf{I} + \phi \mathbfit{u} \mathbfit{u}$. 
Suitable expressions of the equilibrium distributions have been reported before~\cite{swift1996lattice,desplat2001ludwig}. 
For the $f_q^{\rm eq}$, we use
\begin{equation}
	f_q^\text{eq}(\rho, \mathbfit{u}, \boldsymbol{\Pi}) = w_q \left[ \frac{1}{c_s} \rho \mathbfit{u} \cdot H_q^{(1)} + \frac{1}{2 c_s^2} \left( \boldsymbol{\Pi} + \rho \mathbfit{u} \mathbfit{u} \right) : H_q^{(2)} \right]
	\label{eq:eq_dist_fun}
\end{equation}
if $q\neq 0$, and
\begin{equation}
	f_0^\text{eq}(\rho, \mathbfit{u}, \boldsymbol{\Pi}) = \rho - \sum_{q=1}^{Q-1} f_q^\text{eq}.
	\label{eq:eq_dist_fun_0}
\end{equation}
For the $g_q^{\rm eq}$, we use 
\begin{equation}
	g_q^\text{eq} = w_q \left[ \frac{1}{c_s} \phi \mathbfit{u} \cdot H_q^{(1)} + \frac{1}{2 c_s^2} (2 M \vartheta + \phi \mathbfit{u} \mathbfit{u}) : H_q^{(2)} \right],
	\label{eq:phase_eq_dist}
\end{equation}
if $q \neq 0$, and
\begin{equation}
	g_0^\text{eq} = \phi - \sum_{q=1}^{Q-1} g_q^\text{eq}.
		\label{eq:eq_dist_fun_0_g}
\end{equation}
In these expressions, the $w_q$ are weighting factors determined by the geometry of the lattice, $H_q^{(n)} = H^{(n)}(\mathbfit{c}_q)$ is the tensor Hermite polynomial of $n$-th degree, and $c_s = 1/\sqrt{3}$ is a constant that represents the speed of sound~\cite{qian2000dissipative} (see Appendix~\ref{app:D2Q9} for a list of expressions).

Using a Chapman-Enskog expansion, Eqs.~(\ref{eq:lattice-Boltzmann_equation}) and (\ref{eq:lattice-Boltzmann}), together with Eqs.~(\ref{eq:MRT})--(\ref{eq:eq_dist_fun_0_g}), reduce to the Navier-Stokes (\ref{eq:Navier-Stokes}) and Cahn-Hilliard (\ref{eq:Cahn-Hilliard}) equations.
From the expansion, the viscosity, $\mu$, is determined by the coefficients of the collision matrix, $\Lambda_{qr}$~\cite{dHumieres2002mrt} (see Appendix~\ref{app:D2Q9}).

% subsection conservation_of_the_phase_field (end)

\subsection{The electric potential}
\label{subsec:the_electric_potential}

As discussed in \S\ref{sec:diffuse_interface_model}, to model the effect of the electrostatic potential field, it suffices to introduce an algorithm that solves Laplace's equation in the dielectric, whilst keeping the potential to a constant value in the conductor.  

Hence, we take inspiration from the diffusive dynamics which arises from the LBM itself~\cite{ledesma2014lbmevaporation}, 
and introduce a third lattice-Boltzmann equation in the following form,
\begin{equation}
	h_q(\mathbfit{x} + \mathbfit{c}_q, t + 1) = h_q(\mathbfit{x}, t) + \mathcal{C}[h]_q,
	\label{eq:lBe_potential}
\end{equation}
where we use a single-relaxation-time collision operator,
\begin{equation}
	\mathcal{C}[h]_q := - \Lambda [h_q - h_q^\text{eq}](\mathbfit{x}, t),
	\label{eq:MRTh}
\end{equation}
where $\Lambda = 1$. 

This new distribution function is related to the local electric potential, $V$, by the relations 
\begin{equation}
	V = \sum_q h_q,
	\label{eq:potential}
\end{equation}
and
\begin{equation}
	h_q^\text{eq} = w_q V.
	\label{eq:h_eq}
\end{equation}
Eq.~(\ref{eq:h_eq}) offers the advantage of setting the electric potential to a prescribed value, by fixing the right-hand side, 
and thus allows the modelling of a conducting liquid (for which the potential equilibrates to a constant).

We now analyse the long-time, large-lengthscale behaviour of Eqs.~(\ref{eq:lBe_potential})-(\ref{eq:h_eq}). 
First, we express  Eq.~(\ref{eq:lBe_potential}) in terms of the equilibrium distribution, $h^\text{eq}_q$, using Eq.~(\ref{eq:h_eq}).
This is done by writing the collision step as a differential operator acting on $h_q^\text{eq}$ (for details, see Appendix~\ref{app:chapman-enskog}), i.e.,
\begin{equation}
	- \left[h_q - h_q^\text{eq} \right] = (\partial_t + \mathbfit{c}_q \cdot \mathbfit{\nabla}) h_q^\text{eq} - \frac{1}{2} (\partial_t + \mathbfit{c}_q \cdot \mathbfit{\nabla})^2 h_q^\text{eq} + ...
	\label{eq:Picard_lBe}
\end{equation}
Applying the summation operator, $\sum_q$, to Eq.~(\ref{eq:Picard_lBe}), and using Eqs.~(\ref{eq:potential}) and (\ref{eq:h_eq}), we find
\begin{equation}
	0 = \partial_t V - \frac{1}{2} \partial_t^2 V - \frac{1}{2} c_s^2 \nabla^2 V + ...
	\label{eq:V_dynamics}
\end{equation}
where we identify $\epsilon = c_s^2/2$. 
During a relaxation process the first and second terms in Eq.~(\ref{eq:V_dynamics}) will asymptotically vanish, and thus, $V$ will satisfy Eq.~(\ref{eq:harmonic_potential}) at long times.
In the context of electrowetting, one requires that this relaxation is faster than the typical timescales of the hydrodynamic fields, $\mathbfit{u}$ and $\phi$.

To quantify the transient, let us investigate the solutions of Eq.~(\ref{eq:V_dynamics}).
Since the equation is linear, we proceed in the standard way by proposing the \textit{Ansatz} $V = X(\mathbfit{x}) T(t)$~\cite{arfken2013mathematical}.
This leads to the ordinary differential equation for the temporal part,
\begin{equation}
  \label{eq:temp_eq}
  2 \frac{\mathrm{d} T}{\mathrm{d} t} - \frac{\mathrm{d}^2 T}{\mathrm{d} t^2} + c_s^2 K^2 T = 0,
\end{equation}
and a partial differential equation for the spatial part,
\begin{equation}
  \label{eq:spat_eq}
  \nabla^2 X + K^2 X = 0,
\end{equation}
where $K = \text{const.}$, is the eigenvalue that couples the system of equations.

For the temporal part, Eq.~(\ref{eq:temp_eq}), we look for solutions that decay at long times, i.e., 
\begin{equation}
  \label{eq:temp_sol}
  T(t) = \exp \left[(1 - \sqrt{1 + c_s^2 K^2}) \, t \right],
\end{equation}
where the term in brackets is always negative for non-vanishing $K$.

To better understand the rate of decay of the transient, which is controlled by $K$, let us focus on the limiting case of a uniform dielectric phase in a rectangular domain of of size $L_x \times L_y$.
In such a case, Eq.~(\ref{eq:spat_eq}) can be solved analytically~\cite{arfken2013mathematical}, leading to the spectrum of eigenvalues 
\begin{equation}
  \label{eq:possible_K}
  K^2 = \Big( \frac{2 \pi l}{L_x} \Big)^2 + \Big( \frac{2 \pi m}{L_y} \Big)^2,
\end{equation}
where $l$ and $m$ are positive integers.
Let us now define the transient period, $\tau_\text{trans}$, as the characteristic decay time associated to the smallest eigenvalue,
\begin{equation}
  \label{eq:characteristic_time_convergence}
  \tau_\text{trans} := \max \frac{1}{\sqrt{1 + c_s^2 K^2} - 1}, \quad K \neq 0,
\end{equation}
which, for the uniform rectangular domain, is
\begin{equation}
  \label{eq:transient_init}
  \tau_\text{trans} \le \frac{1}{2} \Big( \frac{\max(L_x,\, L_y)}{\pi c_s} \Big)^2.
\end{equation}

The presence of a conductive phase will effectively reduce the domain of Eq.~(\ref{eq:spat_eq}), and thus, will shift the spectrum of $K$ to higher values.
This implies that, Eq.~(\ref{eq:transient_init}) is an upper bound for the transient from arbitrary initial conditions to a steady state solution.

However, if the initial conditions for the electric field are close to a stationary solution, the transient number of iterations required to relax a small perturbation will be much smaller.
For instance, introducing a perturbation of the order of one lattice unit to a stationary solution will lead to $K \approx 2 \pi$.
Hence, from Eq.~(\ref{eq:characteristic_time_convergence}), the transient reduces to
\begin{equation}
  \label{eq:trans_course}
  \tau_\text{trans} \approx \frac{1}{\sqrt{1 + (2 \pi c_s)^2} -1} < 1.
\end{equation}
Such a fast relaxation can be particularly useful, for instance, when the bulk electrostatic potential $V_0$ is varied  quasi statically to explore stationary wetting configurations, were a single iteration might be enough to update the electrostatic field.

% subsection the_electric_potential (end)

\subsection{Simulation setup: initial and boundary conditions}
\label{subsec:boundary_conditions}

We now describe the simulation implementation to model the dynamics in an EWOD setup.
The electric potential and its corresponding distribution function are defined in a simulation box of size $L_x \times L_y$.
The two-phase fluid and corresponding distribution functions are defined in a simulation box of size $L_x \times (L_y - 2d)$, 
where $d$ is a gap used to accommodate for a solid dielectric layer.
This has the purpose of isolating the conductive phase from the bounding electrodes on the finite domain, and thus, to avoid divergences in the electric field.
The permittivity of the solid dielectric is set equal to the permittivity of the dielectric fluid.

The velocity field is set to 
\begin{equation}
\mathbfit{u} (\mathbfit{x},t=0)= 0
\label{eq:u_init_config}
\end{equation} 
everywhere in the simulation domain. 
The phase field, is initialised to 
\begin{equation}
\phi(\mathbfit{x},t=0)=\phi_{\rm i}(\mathbfit{x}),
\label{eq:phi_init_config}
\end{equation} 
which we specify for the specific configurations reported in \S\S~\ref{sec:droplet_spreading} and \ref{sec:dynamics_of_a_thin_dielectric_film}. 
The electric potential is initialised as follows. 
\begin{equation}
  V(\mathbfit{x}, t=0) :=
  \begin{cases}
    V_0, & \text{ if } \phi > 0, \\
    V_0 / 2 & \text{ if } \phi \le 0.
  \end{cases}
  \label{eq:V_init_config}
\end{equation}
At subsequent simulation times, and to smooth out the transition of $V$ from the conductive to the dielectric fluid, we impose the electric potential following the interpolation scheme
\begin{equation}
  V(\mathbfit{x}) = \beta V_0 + (1 - \beta) \sum_q h_q(\mathbfit{x}, t),
  \label{eq:V_value}
\end{equation}
where $\beta$ is an interpolation weight defined as
\begin{equation}
  \beta(\phi(\mathbfit{x})) :=
  \begin{cases}
    1 & \text{ for } \phi > \phi_\text{thr} \\
    \phi / \phi_\text{thr} & \text{ for } 0 < \phi < \phi_\text{thr} \\
    0 & \text{ otherwise},
  \end{cases},
  \label{eq:alpha_interpolation}
\end{equation}
where $\phi_\text{thr} = 0.9$, is a threshold value set to identify the bulk of the conductor.
In this way, the potential is fixed to the prescribed value $V_0$ at the bulk of the conductive phase, whereas it evolves according to Eq.~(\ref{eq:potential}) in the bulk of the dielectric phase.

Using this setup, we found that the electric potential relaxes to a steady state typically after $L_x^2 / 8$ iterations.  
Nonetheless, since transient hydrodynamic flows are slow compared to the speed of sound ($|\mathbfit{u}| \ll c_s$), 
we found that the distribution function $h_q$ could be updated at the same pace as $f_q$ and $g_q$, with only one iteration required to relax the electric potential field.

We impose periodic boundary conditions along the $x$ and $y$ directions,
and fix the solid electrode at the top and bottom boundaries of the simulation domain.
To implement the no-slip boundary condition at the solid surface we use the bounce-back algorithm~\cite{yu2003viscous}.
To implement the wettability of the surface, Eq.~(\ref{eq:Neumann_boundary_condition}), 
we compute the gradient and Laplacian of the phase field at near-boundary nodes using finite differences to then fix the corresponding incoming distribution functions from the solid surface~\cite{briant2004contact-i,desplat2001ludwig}.
Finally, to implement the boundary condition on the voltage, $V_\text{b}$,
we follow a similar approach to that of Ledesma-Aguilar, et al.~\cite{ledesma2014lbmevaporation}. 
We specify the distribution functions streaming from sites on the the solid electrode, of position vector $\mathbfit{x}_\text{b}$, to sites in the fluid near the solid boundary, of position vector $\mathbfit{x}_\text{nb}$, according to
\begin{equation}
 h_{\bar{q}}(\mathbfit{x}_\text{nb}, t) = w_{\bar{q}} \epsilon^{1/2} V_\text{b},
  \label{eq:h_boundary}
\end{equation}
where the indices $\bar{q}$ correspond to the distribution functions that stream away from the boundary.
Specifically, $\bar{q} \in \{ q : \mathbfit{c}_q + \mathbfit{c}_{q'} = 0,\, q \in \Gamma \}$,  
where $\Gamma := \{q : \mathbfit{x}_\text{nb} + \delta \mathbfit{c}_q = \mathbfit{x}_\text{b},\, 0 < \delta < 1 \}$ gives the indices of lattice vectors that stream towards the electrode.

% subsection boundary_conditions (end)

\begin{table}[b]
  \centering  \small
  \begin{tabular}{l c c c}
    Simulation parameter & Symbol & Value in \S\ref{subsec:spreading_sims} & Value in \S\ref{subsec:dielectric_film_sims} \\
    \hline
    Simulation box & $L_x \times L_y$ & $512 \times 288$ & $[418, 1256] \times 84$ \\
    Surface tension & $\gamma$ & $6 \times 10^{-3}$ & $8 \times 10^{-3}$ \\
    Interface width & $\ell$ & $4$ & $5$ \\
    Contact angle &  $\theta_0$ & $[160^\circ, 120^\circ]$ & $180^\circ$ \\
    Density & $\rho$ & $1$ & $1$ \\
    Dynamic viscosity & $\mu_\text{c}$, $\mu_\text{d}$ & $1/6$, $1/6$ & $1/600$, $1/3$ \\
    Mobility & $M$ & $1/10$ & $1/10$ \\
    Permittivity & $\varepsilon$ & 1/6 & 1/6\\
    Dielectric thickness & $d$ & $2$ & $2$ \\
    Initial config. & & $R_0 = 128$ & $a = 1$, $H_0 = 20$ \\
    \hline
  \end{tabular}
  \caption{Parameters for the simulations of the spreading of a droplet and the dielectric film dynamics.}
  \label{tab:sim_params}
\end{table}

% section lattice_boltzmann_simulations (end)

\section{Electrowetting of a droplet}
\label{sec:droplet_spreading}

In this section we validate the lattice-Boltzmann algorithm by studying the electrowetting-driven spreading of a droplet in an EWOD setup. 
We start by reviewing the Young-Lippmann classical theory of electrowetting~\cite{lippmann1875relations,mugele2005electrowetting}, before comparing to our simulation results.

\subsection{Review of the Young-Lipmann Theory}
\label{sec:spreading_theo}

\begin{figure}[t]
	\centering
		\includegraphics[width=\columnwidth]{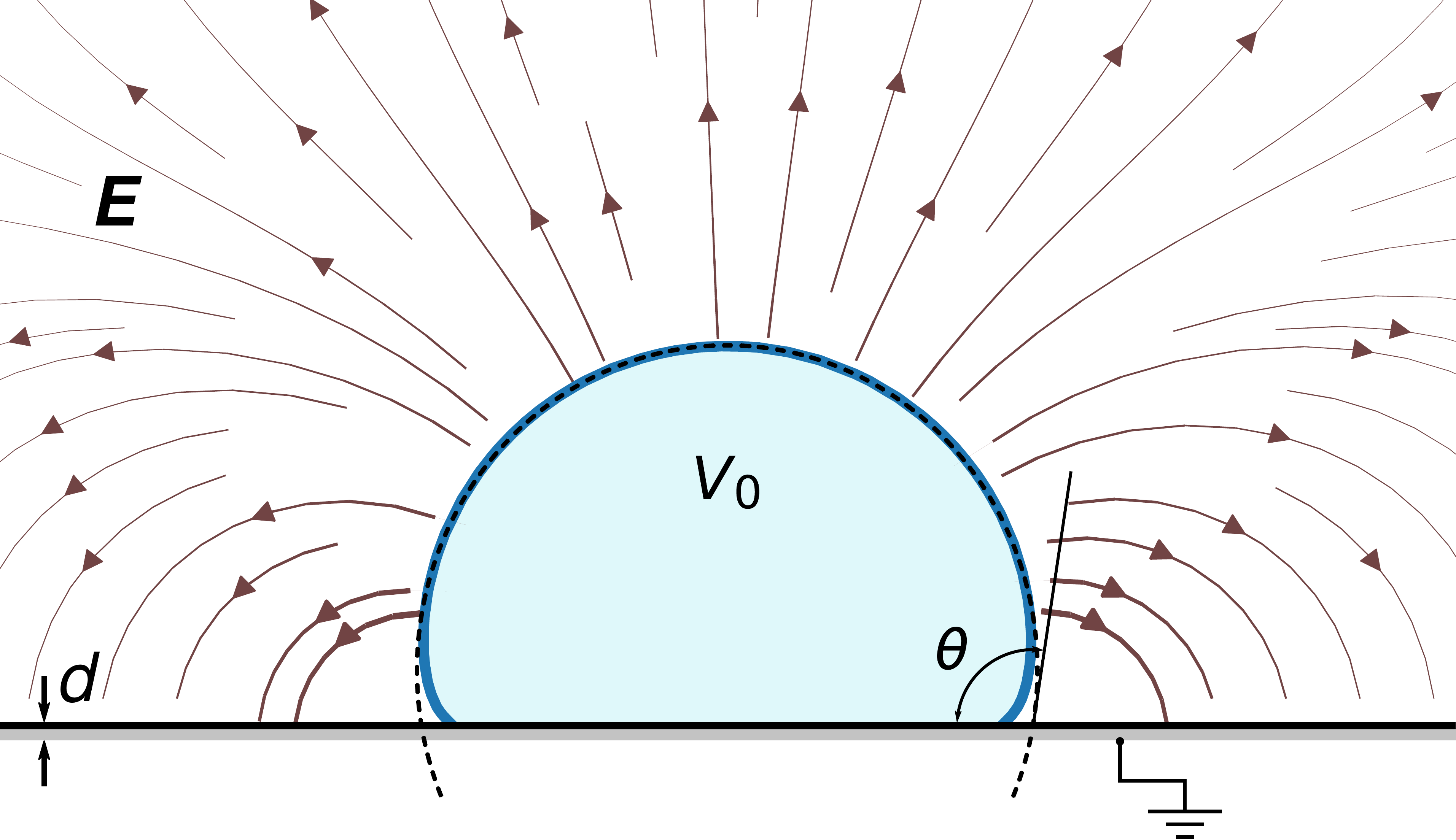}
    \caption{(Colour online) 2D LB simulations of a droplet in an EWOD set-up.
    A droplet of conducting liquid sits on top of a dielectric solid of thickness $d$.
    The droplet is set to an electric potential $V_0$ and, on the other side of the dielectric surface, the electric potential is set to zero.
    The dielectric fluid surrounds the droplet where the electric field, $\mathbfit{E}$ is shown by the stream plot.
    The dashed line corresponds to the best fitting circle to the cap and intersects the solid surface at the apparent contact angle $\theta$.
    }
	\label{fig:2D_configurations}
\end{figure}

Consider a droplet of a conductive liquid in an EWOD setup as shown in Fig.~\ref{fig:2D_configurations}.
As the potential difference applied between the droplet and the electrode is increased, the electric charges begin to gather at the interface of a conductive liquid with a higher density near the grounded electrode.
This configuration corresponds to a capacitor.
Therefore, and neglecting the charges that accumulate on the opposite side of the solid dielectric, 
the electrostatic energy, per unit surface area of the electrode, is $c V_0^2 / 2$, where $c$ is the capacitance per unit area of the configuration~\cite{landau2013electrodynamics}. 
Because the droplet's surface is compliant, the electrostatic force leads to a spreading of the liquid on the solid electrode. 

The equilibrium configuration of the droplet will be determined by the balance between the work done by the electric field against the increase in surface energy. 
Mechanically, an infinitesimal radial displacement of the contact line, $\mathrm{d}R$, results in a net radial force on the interface of the droplet. 
Hence, mechanical equilibrium is achieved when 
\begin{equation}
  0 = (\gamma_\text{sd} - \gamma_\text{sc} - \gamma \cos \theta + c V_0^2 / 2) \mathrm{d}R.
  \label{eq:pre_lippmann}
\end{equation}
Using Eq.~(\ref{eq:young-dupre}) and dividing by $\gamma$, Eq.~(\ref{eq:pre_lippmann}) results in the Young-Lippmann relation~\cite{mugele2005electrowetting},
\begin{equation}
  \cos \theta(V_0) = \cos \theta_0 + \eta,
	\label{eq:young-lippmann}
\end{equation}
where 
\begin{equation}
	\eta := \frac{c V_0^2}{2 \gamma}
\end{equation}
is the electrowetting number. 

Therefore, the contact angle of a droplet is reduced with increasing applied voltage.
Experimentally, Young-Lipmann's result has been verified over a range of voltages. However, it has also been observed that at high voltages the contact angle reaches a saturation value, beyond which the theory is no longer valid~\cite{quinn2005contact,peykov2000electrowetting}.

% subsection spreading_theo (end)

\subsection{Lattice-Boltzmann simulations}
\label{subsec:spreading_sims}

\begin{figure}[t!]
	\centering
	\includegraphics[width=\columnwidth]{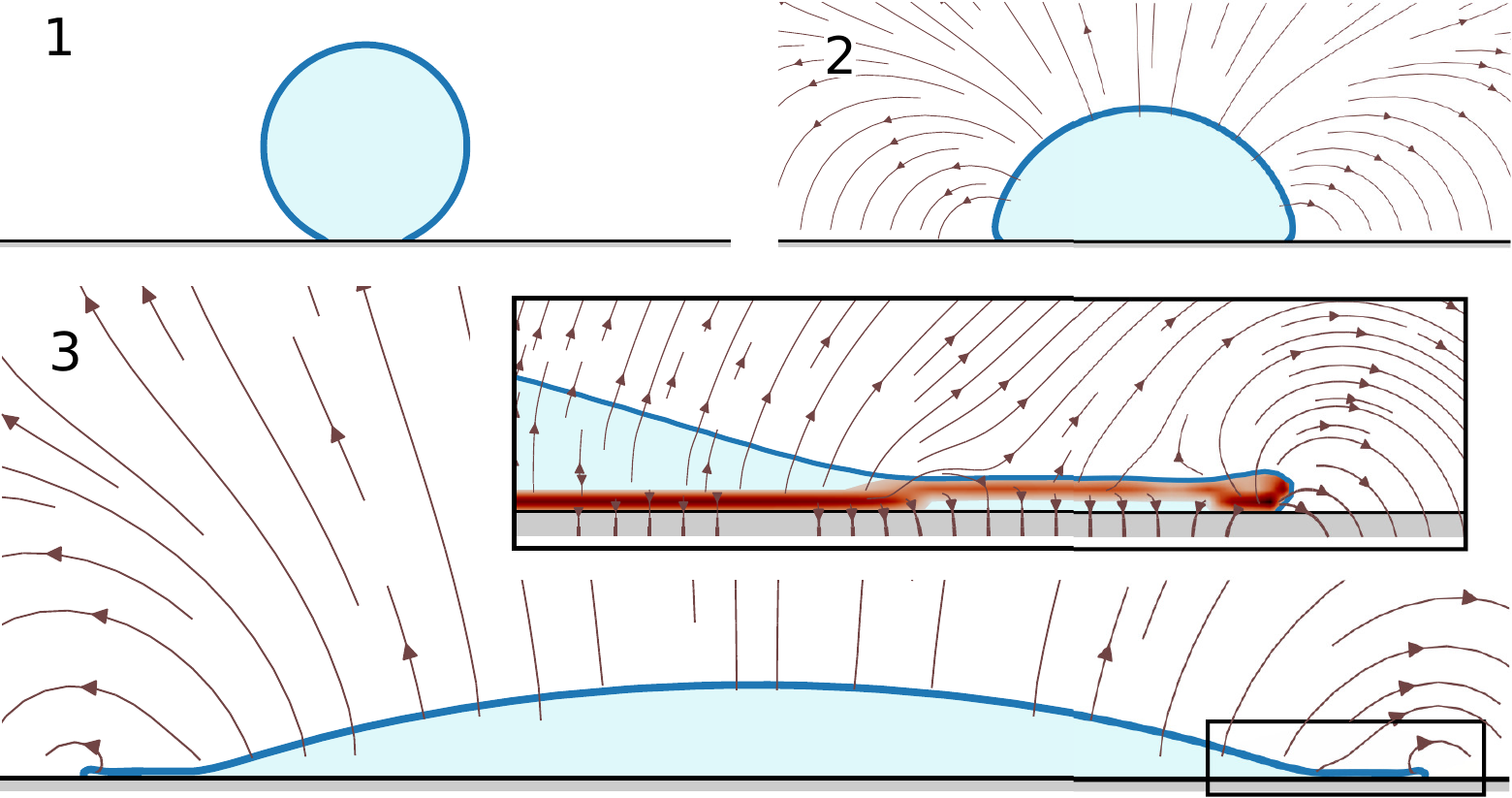} 
	(\textit{a})
  \vspace{4mm}
  
  \includegraphics[width=\columnwidth]{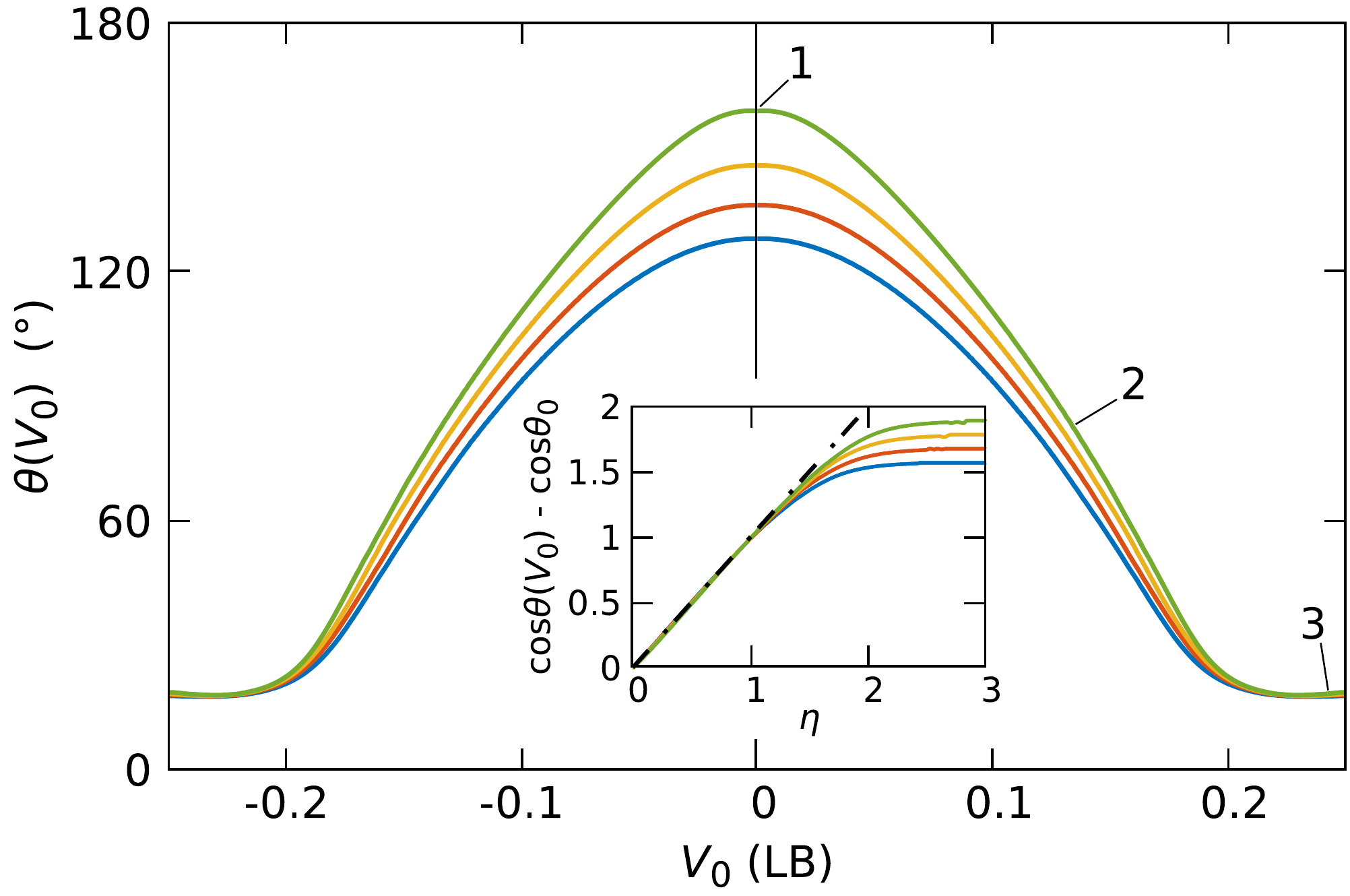}\\
  (\textit{b})
	\caption{Simulations of droplet spreading using an EWOD setup. 
    (\textit{a}) Stationary droplet configurations at different applied voltage, $V_0$.
    At $V_0 = 0$, the shape of a droplet is circular and intersects the solid dielectric at the equilibrium contact angle $\theta_0$.
    For $|V_0|>0$, the shape of the droplet close to the solid wall is distorted by the electric field, leading to an apparent contact angle, $\theta(V_0)$. 
    At high applied voltages, the droplet reaches a limiting configuration, where the main drop develops a lip that spreads away from its centre.
    The region around the lip shows strong fringe fields (inset) and the charge density (dark-red colour map).
    (\textit{b}) Variation of the contact angle in response to the electric potential, $V_0$, in lattice-Boltzmann units.
    The curves show a monotonic decrease in the contact angle with the increasing magnitude of the potential.
    The inset shows the expected universal collapse as a function of the electrowetting number, $\eta$, predicted by the Young-Lippmann relation (dotted-dashed line) at low electrowetting numbers and a later saturation.
  }
	\label{fig:contact_angle_response}
\end{figure}

The initial configuration of the system consists of a circular droplet of the conducting liquid suspended in the dielectric fluid. 
We impose the initial conditions in the simulations using Eqs.~(\ref{eq:u_init_config}), (\ref{eq:phi_init_config}) and (\ref{eq:V_init_config}); the initial phase field reads
\begin{align}
  \phi_{\rm i}(\mathbfit{x}) & = \tanh \left(\frac{R_0 - |\mathbfit{x} - \mathbfit{X}_0|}{\sqrt{2}\ell} \right), 
  \end{align}
where $\mathbfit{X}_0 = (L_x / 2, R_0)$, is the initial position of the centre of the droplet, and $R_0$ its initial radius. 
The rest of the simulation parameters are summarised in Table~\ref{tab:sim_params}.

We first set the potential within the conducting droplet to $V_0 = 0$ and allow the system to relax for $2 \times 10^5$ iterations.
As the droplet relaxes, it spreads on the surface and acquires a circular-cap shape intersecting the surface with the expected equilibrium contact angle,  $\theta_0$, predicted by 
Eq.~(\ref{eq:wetting_potential}).
Then, we increase the voltage by an amount $0.01 \sqrt{2 \gamma d / \varepsilon}$ and allow the system to relax for a further $10^4$ iterations.
Once the relaxation has elapsed, the stationary configuration is recorded. 
The increment in the applied voltage is repeated until a maximum voltage $V_0 = 3 \sqrt{2 \gamma d / \varepsilon}$ is reached.

Fig.~\ref{fig:2D_configurations} shows a typical equilibrium configuration of the droplet subject to a non-zero potential. 
The upper part of the droplet conserves a circular shape that, extrapolated, intersects the surface at an apparent contact angle $\theta(V_0)$.
However, near the solid surface, the inclination of the interface is closer to the prescribed equilibrium contact angle.
% 
%Buehrle \textit{et al.}.
%
% {\color{blue} Has this been observed in simulations/experiments before?}
%
As shown in Fig.~\ref{fig:contact_angle_response}b, the apparent contact angle decreases with increasing $|V_0|$. 
Note that reversing the polarity of the applied voltage leads to the same decrease in the apparent angle; 
this is expected, since Eq.~(\ref{eq:electrical_force}) is invariant upon an inversion of the polarity of the electric potential ($V \rightarrow - V$).
Therefore, the simulations capture the competition between electrical and capillary forces, as has been reported previously in experimental observations~\cite{buehrle2003profiles}.

Next, we carried out simulations to measure $\theta(V_0)$ for different values of the equilibrium contact angle, $\theta_0$. 
As shown in Fig.~\ref{fig:contact_angle_response}b, the $\theta(V_0)$ curves follow the same trend, with only a shift of the maximum to a value imposed by $\theta_0$. 
As shown in the inset, a plot of $\cos \theta(V_0) - \cos \theta_0$ shows a linear dependence on $\eta$, 
which is in agreement with the theoretical prediction of Eq.~(\ref{eq:young-lippmann}).  
Fitting the simulation data to a straight line gives $c \approx 0.66$.

As the voltage in the droplet is increased, the apparent contact angle reaches a saturation value $\theta \approx 18.43^\circ$.
The saturation effect was found to be independent of the wettability of the surface, and begins to occur when the droplet reaches $\theta \sim 50^\circ$.
From the simulations, we observe that at the onset of saturation the droplet develops two distinct regions.
Close to its centre, the capillary forces smooth out the shape of the interface, which remains circular. 
However, the region close to the edge is subject to strong fringe fields, and deforms to take the shape of a `lip', spreading away from the main drop (see panel 4 in Fig.~\ref{fig:contact_angle_response}(a)).
The result is that the bulk profile retains a limiting shape, characterised by the saturation contact angle, while an increase in the voltage results in a further growth of the edge lip. 
%

% subsection spreading_sims

% section droplet_spreading (end)

\section{Dynamics of a thin dielectric film}
\label{sec:dynamics_of_a_thin_dielectric_film}

\begin{figure}[t!]
  \centering
  \includegraphics[width=0.7\columnwidth]{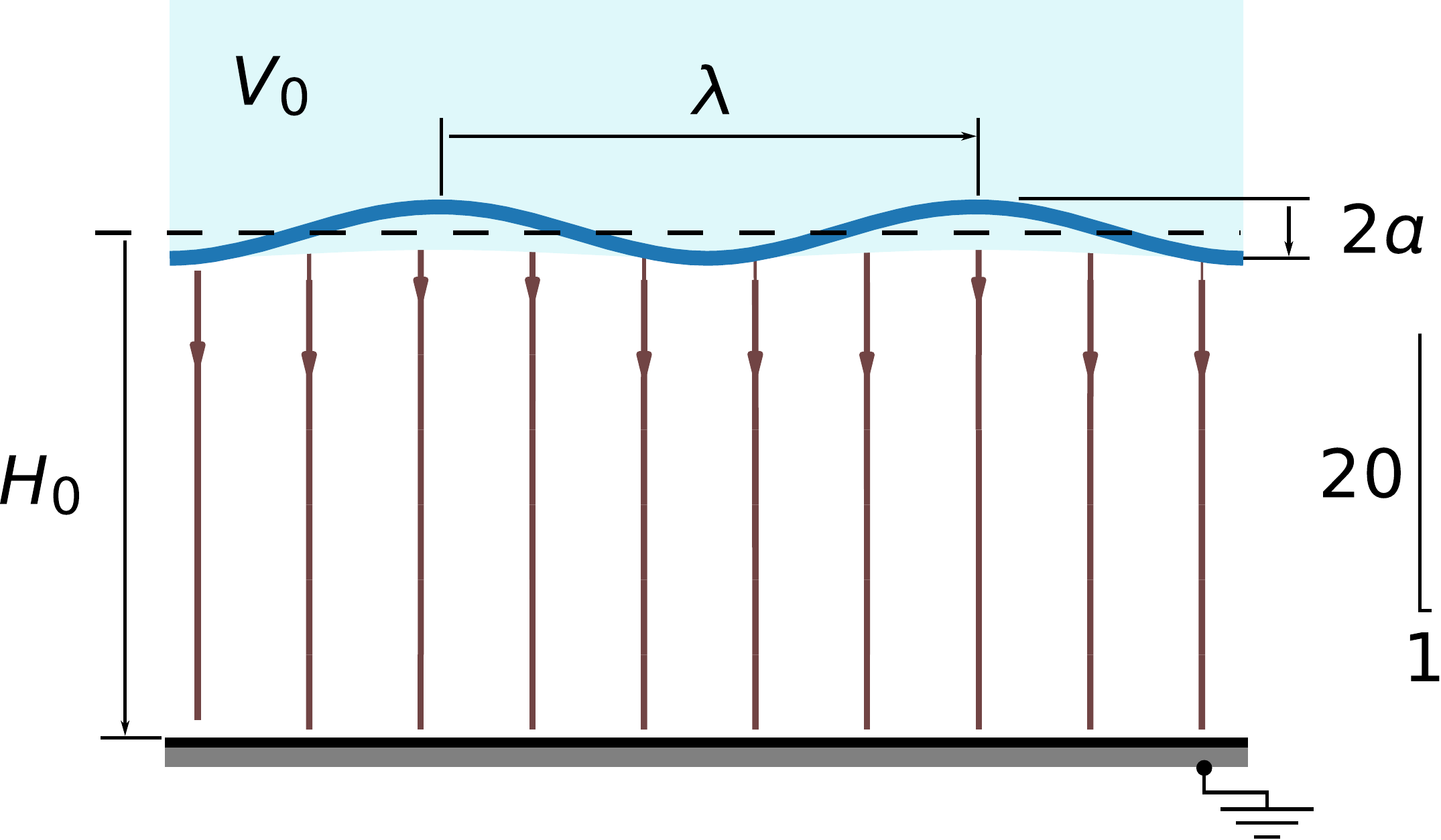} \\ (\textit{a}) \\  \vspace{2mm}
	\begin{tabular}{c c}
  \includegraphics[width=0.45\columnwidth]{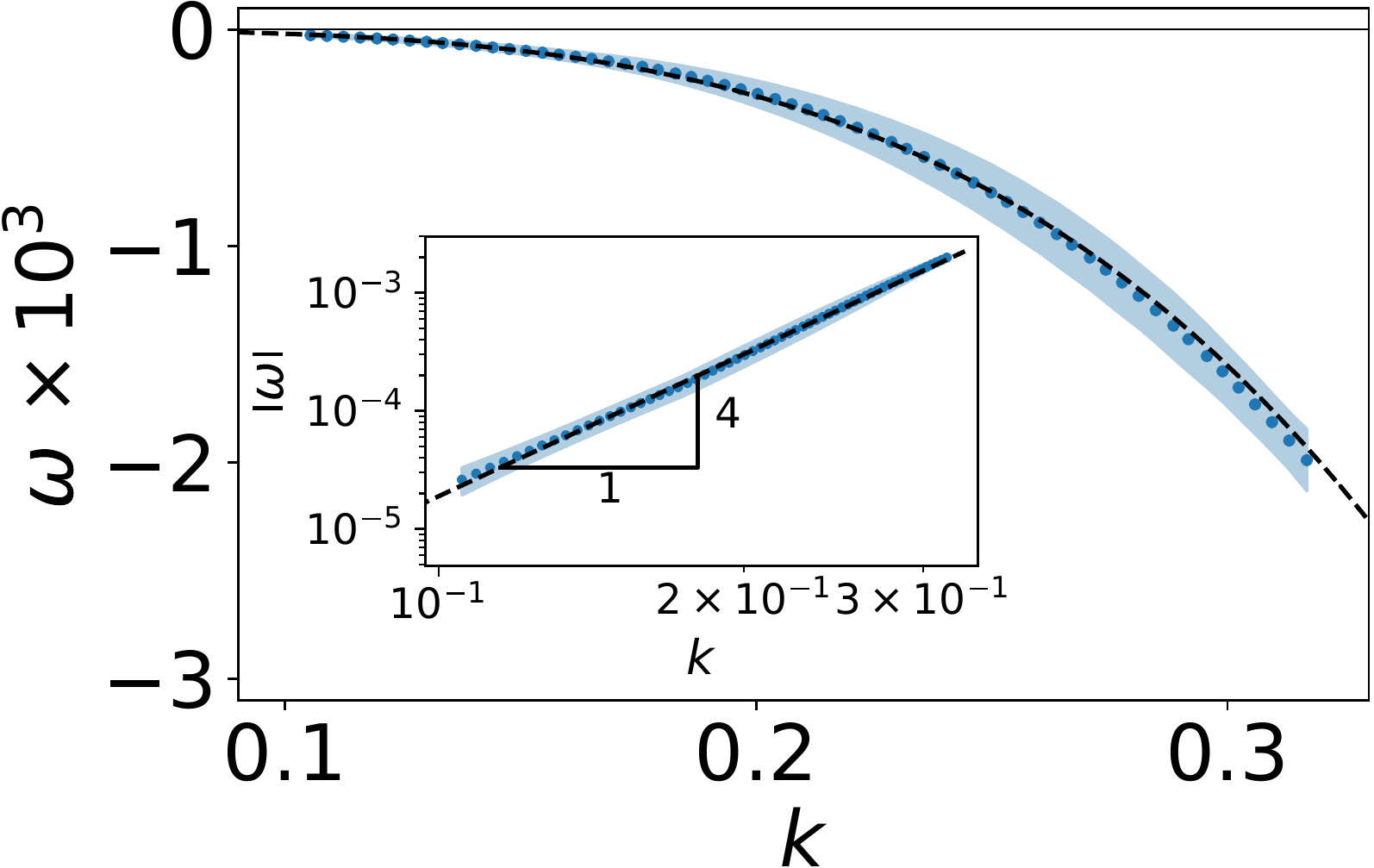} &
  \includegraphics[width=0.45\columnwidth]{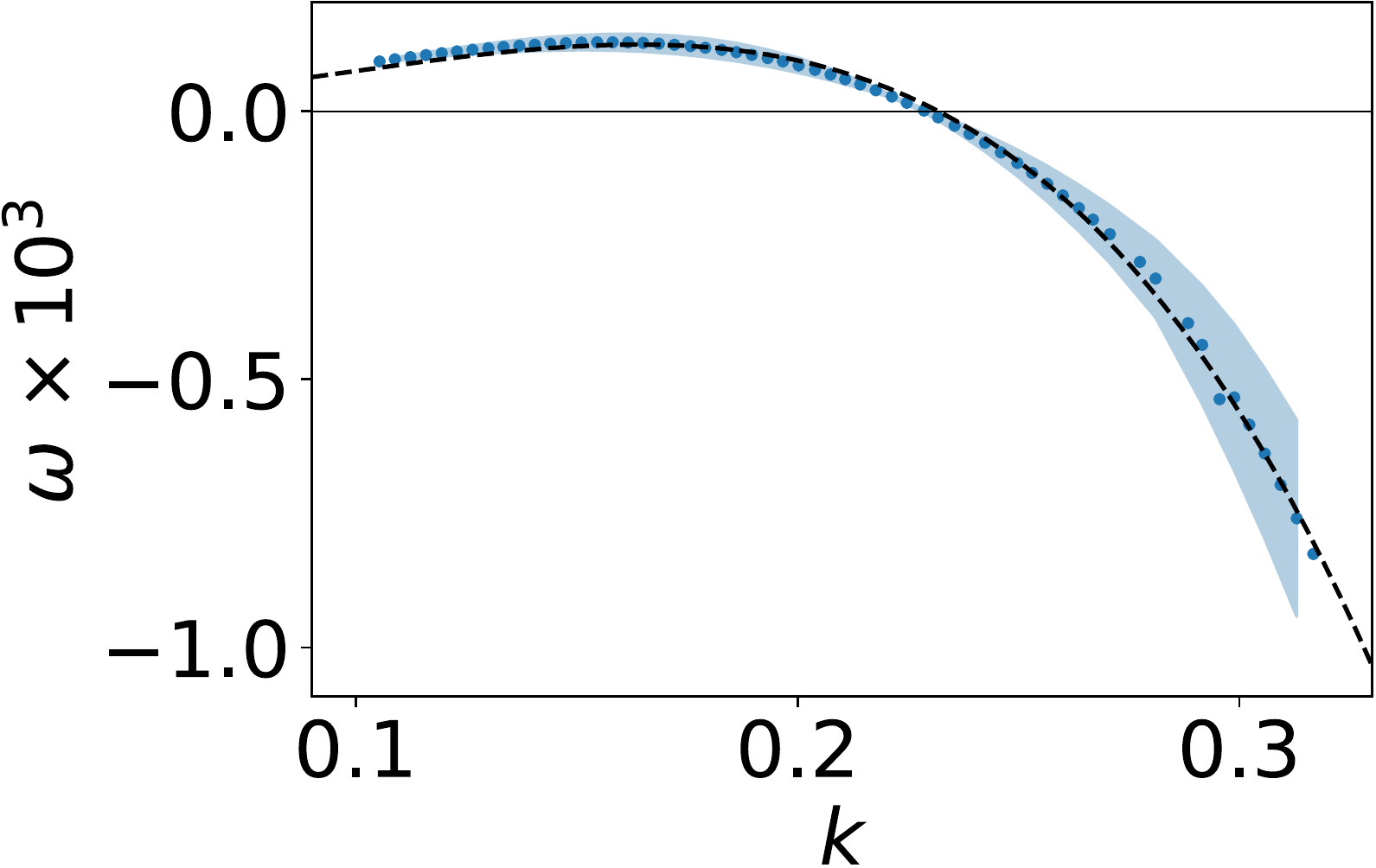} \\
    (\textit{b}) & (\textit{c})
	\end{tabular} \\ \vspace{2mm}
	\includegraphics[width=\columnwidth]{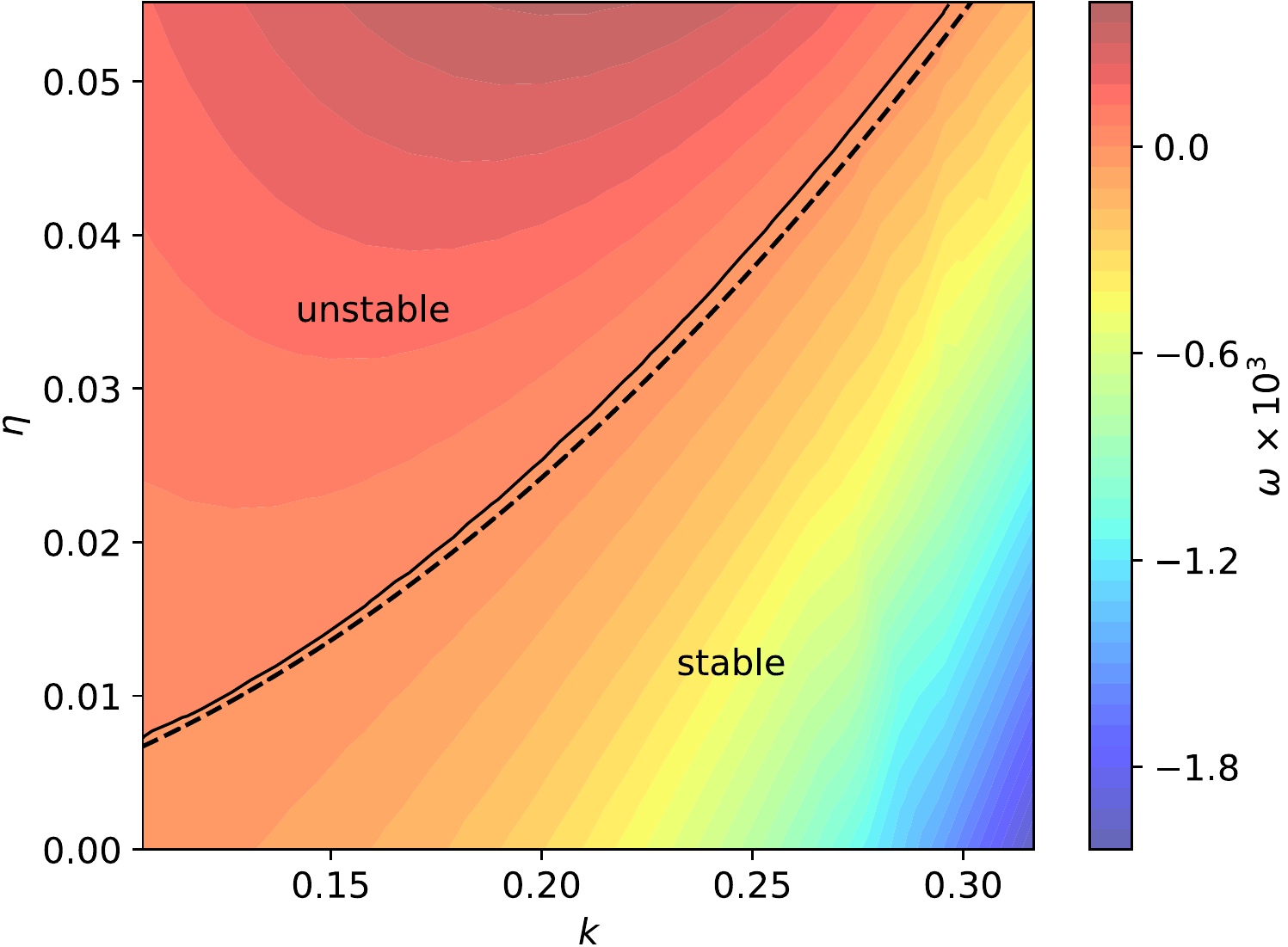} \\ (\textit{d})  
  \caption[Thin dielectric film simulation]{
    (Colour online) Simulation results of the linear stability of a thin dielectric fluid film in an EWOD setup. 
    (\textit{a}) Close-up of the interface configuration. 
    The conducting liquid (light blue region) is kept at a constant voltage $V_0$, whilst the solid electrode (grey rectangle) remains grounded. 
    The thin dielectric fluid film (white region), of initial average thickness $H_0$, is subject to a sinusoidal perturbation of amplitude $a$ and wavelength $\lambda$. 
    Direct contact between the dielectric fluid and the solid electrode is prevented by a thin dielectric film (black line). 
    The stream lines depict the electric field.
    (b) and (c): Dispersion relations for $\eta=0$ and $\eta = 0.03$, respectively. 
    The solid symbols correspond to the simulation results. 
    The dashed lines correspond to a fit to the analytical model. 
    The shaded envelopes represent the error from the curve fitting analysis.
    The inset shows the expected $|\omega|\sim k^4$ scaling predicted by the linear theory. 
    (\textit{d}) Colour map of the growth rate as function of $\eta$ and $k$. 
    The solid line corresponds to the separatrix calculated from the simulation results using linear interpolation. 
    The dashed line corresponds to the theoretical prediction. 
  }
  \label{fig:film_instability_sim}
\end{figure}

In this section we illustrate the applicability of the lattice-Boltzmann algorithm to resolve the dynamics of electrowetting liquids. 
Specifically, we study the stability of a thin dielectric film confined between a solid charged wall and a conductive liquid layer. 
This problem is relevant in many electrowetting setups, 
where the spreading conductive liquid often entraps a thin film of dielectric fluid. 
%
%This is more noticeable for high viscosity of the dielectric phase.
%
As the dielectric film becomes thinner, it breaks up into small droplets~\cite{staicu2006entrapment}. 

We start by formulating the problem analytically, which yields a prediction of the stability of the film in the linear regime. 
We then report simulation results which we validate against this prediction, 
and extend our study to report results of the dynamics of the film at long times, including the regime of film breakup and droplet formation. 

\subsection{Linear-stability theory}
\label{subsec:dielectric_film_theo}
We consider a thin, two-dimensional dielectric film of local thickness $H(x,t)$. 
The film lies on top of a conducting solid electrode, located at $y=-d$ which is coated with a thin dielectric solid layer of thickness $d$. 
At its top, the film is covered by a layer of conducting liquid of negligible viscosity. 

To model the dynamics of the thin dielectric layer in the presence of an electric field, we use the lubrication equation~\cite{oron1997evolution}, 
\begin{equation}
    \partial_t H - \partial_x \left( \frac{H^3}{3 \mu_{\rm d}}\partial_x p_\text{film} \right) = 0.
  \label{eq:thin-film-eqn}
\end{equation}
As shown by Eq.~(\ref{eq:thin-film-eqn}), the dynamics is driven by variations in the pressure within the film, $p_\text{film}$. 
This is composed of a capillary contribution, $2\gamma \kappa$, and by a contribution due to the electric stresses on the dielectric fluid, $-\frac{1}{2}\partial_H c V_0^2.$ 
For a gently curved interface, $\kappa \approx -\frac{1}{2}\partial^2_xH $. 
Hence, 
\begin{equation}
  p_\text{film} = - \gamma \partial_x^2H - \frac{1}{2} (\partial_H c) V_0^2.
  \label{eq:pressure_film}
\end{equation}
where we assume that the capacitance $c$ for a dielectric film in contact with the dielectric solid layer is given by 
\begin{equation}
  c = \frac{\varepsilon}{H + d}.
  \label{eq:cap_dielectric_film}
\end{equation}

We now study the stability the dielectric film by analysing Eq.~(\ref{eq:thin-film-eqn}) using a perturbative approach.
Let us consider the initial sinusoidal interface profile
\begin{equation}
  H(x, t) = H_0 + a \cos( 2 \pi x / \lambda ) \exp(t / \tau),
  \label{eq:film_thickness}
\end{equation}
where $H_0$ is the average height of the film, $a$ is the amplitude of the perturbation, $\lambda$ the wavelength and $\tau$ is the characteristic growth time.

Substituting Eq.~(\ref{eq:film_thickness}) into Eq.~(\ref{eq:thin-film-eqn}), and assuming $a \ll H_0$ gives the dispersion relation
\begin{equation}
  \omega = \frac{1}{3} k^2 \left[ 2 \eta \left(\frac{H_0}{H_0 + d}\right)^2 - k^2 \right],
  \label{eq:dispersion_relation}
\end{equation}
where $\omega := \mu_{\rm d} H_0 / \gamma \tau$ is the dimensionless growth rate, and $k := 2 \pi H_0 / \lambda$ is the dimensionless wave number. 

The first term in Eq.~(\ref{eq:dispersion_relation}) corresponds to the destabilising effect of the electric field, which dominates for long-wavelength perturbations. 
This competes against the stabilising effect of surface tension, which dominates for short wavelengths. 
Setting $\omega=0$, corresponding to the onset of instability, gives the separatrix 
\begin{equation}
  \eta = \frac{1}{2} \left( \frac{H_0 + d}{H_0} \right)^2 k^2,
  \label{eq:separatrix}
\end{equation}
which gives the minimum electrowetting number for which a perturbation of given wave number leads to instability. 

% subsection dielectric_film_theo

\subsection{LB simulations}
\label{subsec:dielectric_film_sims}
We impose the initial conditions in the simulations using Eqs.~(\ref{eq:u_init_config}), (\ref{eq:phi_init_config}) and (\ref{eq:V_init_config}); 
we introduce an initial perturbation to the interface between the conductive and dielectric fluids by imposing the phase-field profile 
\begin{align}
 \phi_\text{i}(\mathbfit{x}) & = \tanh \left( \frac{y - H(x,0)}{\sqrt{2} \ell} \right), \label{eq:init_phi_dielectric_film}
\end{align}
with corresponds to a sinusoidal perturbation of amplitude $a=1$ and wavelength $\lambda = L_x$.
The rest of the simulation parameters are reported in the last column of Table~\ref{tab:sim_params}.
To allow the thermodynamic relaxation of the phase field from the initial conditions, we let the simulations run for $10^3$ iterations, which we disregard. 

Fig.~\ref{fig:film_instability_sim}(\textit{a}) shows a typical instantaneous configuration of the film after the transient has elapsed. 
Henceforth, we track the evolution of the fluid-fluid interface, whose location we take as the level curve $\phi(x,y) = 0$. 
Once the location of the interface is determined, 
the amplitude of the perturbation is found by fitting the instantaneous level curves to the sinusoidal function $y(x)=c_0 + c_1 \cos(2 \pi x / L_x)$, where $c_0$ and $c_1$ are fitting parameters. 
We then fit the measured amplitude data, $c_1(t)$, to the exponential function $A(t) = c_2 \exp(t / c_3)$, where $c_3$ gives the characteristic growth time.
To obtain the dependence of the dispersion relation, for a given electrowetting number, we repeat the simulation by varying the system length, $L_x$ (see Table~\ref{tab:sim_params}). 

Figs.~\ref{fig:film_instability_sim}(\textit{b}) and \ref{fig:film_instability_sim}(\textit{c}) show the dispersion relations obtained from the simulations for $\eta =0 $ and $\eta=0.03$. 
The data in the figures is reported in the dimensionless units of Eq.~(\ref{eq:dispersion_relation}), where $\mu_{\rm d}$, $\gamma$, $H_0$ are fixed using the values reported in Table~\ref{tab:sim_params}.  
For $\eta =0$, we observe the expected power-law decay, $\omega \propto -k^4$, predicted by the linear stability analysis. 
For $\eta =0.03$, the dispersion relation shows a range of unstable wave numbers. 
In both cases, we find a quantitative agreement with Eq.~(\ref{eq:dispersion_relation}), which is superimposed to the simulation data as a dashed line. 

We measured the growth rate of the perturbation for  $21\times21$ points in  the $\eta$--$k$ space.
Fig.~\ref{fig:film_instability_sim}(\textit{d}) shows the simulation results, which we present as a contour plot of $\omega$ vs $\eta$ and $k$. 
The separatrix, corresponding to the curve $\omega(k,\eta)=0$, was estimated from the data using bilinear interpolation (solid line in the figure). 
Overall, there is a good agreement with Eq.~(\ref{eq:separatrix}) (shown as a dashed line). 
We attribute the small discrepancy between the theory and the simulation results to the charge distribution at the diffuse fluid-fluid interface, 
which is dispersed in a region of the order of the interface thickness $\ell$. 
This effect would then alter the capacitance of the dielectric film. 
Indeed, by fitting the separatrix obtained from the simulations to Eq.~(\ref{eq:separatrix}), we obtain an effective value for $H_0$, which is displaced by a small amount ($\sim 0.08 \ell$) into the bulk of the conductive phase.

\begin{figure}[t!]
  \centering

  \includegraphics[width=\columnwidth]{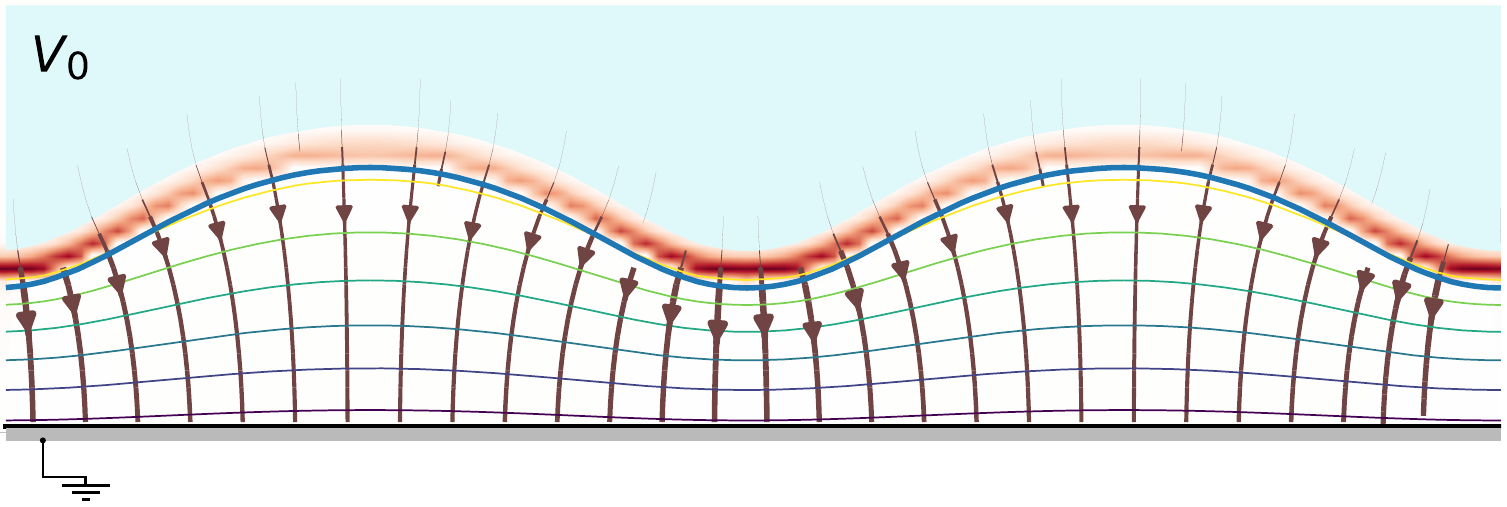} \\ (\textit{a})
  \includegraphics[width=\columnwidth]{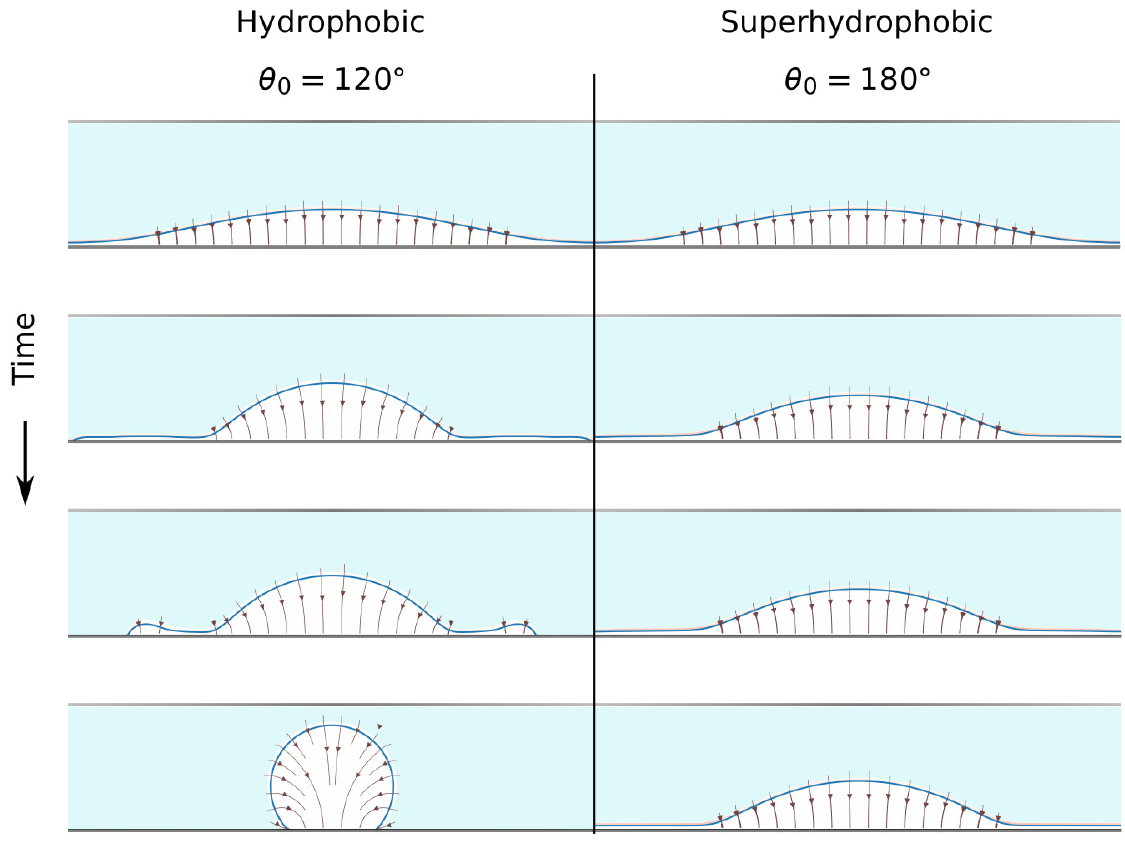} \\ (\textit{b})
  % \hspace{0.05\columnwidth} (\textit{b}) \hspace{0.4\columnwidth} (\textit{c})
  \caption[Film breakup into droplets.]{
    (Colour online) Entrapment and break-up of unstable dielectric films. 
    (\textit{a}) Instantaneous configuration of an unstable dielectric film at large perturbation amplitudes. The initial simulation parameters are $\lambda = 60$, $H_0 = 20$ and $\eta = 1.1$.  
    The charge distribution, shown in dark red, is highest in the regions closer to the solid electrode, and the equipotential curves, perpendicular to the electric field, increase in density.
    (\textit{b}) Long-time evolution of the dielectric film for $\lambda = 512$, $H_0 = 20$, and $\eta = 0.1$; and $\theta_0=120^\circ$ (left) and $\theta_0=180^\circ$ (right). 
    On a wettable surface, the dielectric fluid breaks into isolated films that dewet to form droplets.  
    On a non-wettable surface the wetting potential prevents the breakup of the film, leading to its entrapment. 
  }
  \label{fig:film_breakup}
\end{figure}

We now turn our attention to the growth of the perturbation at long times, when $a/H_0\sim 1$. 
This regime, which is not accessible by the linear theory, is revealed in detail by the simulations. 
As shown in Fig.~\ref{fig:film_breakup}(\textit{a}), at large perturbation amplitudes inhomogeneities in the electric field become apparent.
The simulations capture the increase in charge density in regions where the interface curvature is higher~\cite{jackson1999classical}. 
This effect leads to a stronger electrostatic attraction in regions of the interface which lie closer to the solid electrode.
As a result, the perturbation grows faster than predicted by the linear theory, and the interface is deformed to an asymmetric shape. 

At longer times, the troughs of the perturbation approach the solid surface. 
In this regime, we found that the wettability of the solid has a strong effect on the dynamics. 
For $\theta_0<180^\circ$ the fluid-fluid interface touches the solid surface, breaking the film into droplets. 
The subsequent dynamics of the fluid-fluid interface is similar to the dewetting dynamics observed by Edwards, et al.~\cite{edwards2016notspreading}:
the retracting edges collect fluid to form dewetting rims, which eventually merge to form a single circular droplet (see Fig.~\ref{fig:film_breakup}(\textit{b})).
For $\theta_0=180^\circ$, the conducting fluid cannot wet the surface and, hence, the dielectric film does not break up. 
Therefore, the film takes the shape of a series of `bumps'  which remain connected by a thin film (of a thickness set by the range of the wetting potential in the simulations).
This situation is reminiscent of the oil entrapment regime reported by Saticu et al.~\cite{staicu2006entrapment}, 
who used an EWOD setup to spread water droplets immersed in silicone oil on Teflon-coated electrodes. 
% 

% subsection dielectric_film_sims (end)

% section dynamics_of_a_thin_dielectric_film (end)

\section{Conclusions}
\label{sec:conclusions}
We have presented a lattice-Boltzmann algorithm capable of solving the coupled hydrodynamics and electrostatics equations of motion of a two-phase fluid. 
The main advantage of our model is its ability to solve the electrostatics equations within the lattice-Boltzmann algorithm itself, eliminating the need for concurrent methods, such as finite differences or finite element methods, to model the electric field. 

We have validated our algorithm by presenting numerical simulations of the electrowetting of a droplet in an Electro Wetting On Dielectric (EWOD) setup.
Our results reproduce the dependence of the apparent contact angle of the droplet on the applied voltage predicted by the Young-Lippman theory. 
We also observe a saturation of the contact angle at high voltages. 
The saturation of the contact angle has been reported in experiments, and remains an open question in the field of electrowetting.  
In the simulations, the effect is linked to a saturation of the interface curvature, which triggers the formation of a `lip' at the droplet's edge.
Such a balance between the electric and capillary stresses in the simulations might explain the saturation effect observed in experiments, but further experimental evidence is needed to reach a conclusion in this regard. 

We have also used our algorithm to study the stability and dynamics of a thin dielectric film in an EWOD setup.  
For small perturbations, our simulations results agree well with the prediction of lubrication theory. 
Beyond this small-perturbation regime accessible by theory, we studied the long-time dynamics of the film.  
Our simulations show that as the film is destabilised and the interface approaches the solid surface. 
On wettable surfaces, the film breaks up and forms droplets that dewet from the surface. 
On non-wettable surfaces, we observe the entrapment of the dielectric film and the stabilisation of mound-shaped structures. 
%

% section conclusions (end)

% \section*{Conflicts of interest}

% There are no conflicts of interest to declare.

\section*{Acknowledgements}

The authors acknowledge support from EPSRC First-Grant Scheme (No. EP/P024408/1) and from EPSRC's UK Consortium on Mesoscale Engineering Sciences (No. EP/R029598/1). 

\section*{Appendix}
\label{sec:appendix}

\appendix
\section{Parameters of the D2Q9 model}
\label{app:D2Q9}

The set of velocities, $\{\mathbfit{c}_q\}_{q=0}^{Q-1}$, for the D2Q9 LBM is:
\begin{tabular}{l l l}
  $\mathbfit{c}_0 = (0,\, 0), \qquad \qquad$ & $\mathbfit{c}_1 = (1,\, 0), \qquad \qquad$ & $\mathbfit{c}_2 = (0,\, 1),$ \\
  $\mathbfit{c}_3 = (-1,\, 0),$ & $\mathbfit{c}_4 = (0,\, -1),$ & $\mathbfit{c}_5 = (1,\, 1), $ \\
  $\mathbfit{c}_6 = (-1,\, 1),$ & $\mathbfit{c}_7 = (-1,\, -1), \quad \text{ and } $ & $\mathbfit{c}_8 = (1,\, -1).$
\end{tabular}
The set of weights are: $w_0 = 4/9$, $w_{1 \text{--} 4} = 1/9$, and $w_{5 \text{--} 8} = 1/36$.

The tensor Hermite Polynomials are defined as~\cite{kruger2016lattice},
\begin{equation}
  \label{eq:hermite_def}
  H^{(n)}(\mathbfit{x}) := (-1)^n e^{|\mathbfit{x}|^2/2} \nabla^n e^{-|\mathbfit{x}|^2/2}.
\end{equation}
Explicitly, the first are,
\begin{align}
  H^{(0)}(\mathbfit{c}_q/c_s) & = 1, \\
  H^{(1)}(\mathbfit{c}_q/c_s) & = \mathbfit{c}_q / c_s, \quad \text{and} \\
  H^{(2)}(\mathbfit{c}_q/c_s) & = \mathbfit{c}_q \mathbfit{c}_q / c_s^2 - \mathbf{I}.
\end{align}

The collision matrix, $\Lambda_{kn}$ is defined as,
\begin{equation}
  \Lambda_{kn} := \sum_{l=0}^{Q-1} \sum_{m=0}^{Q-1} (M^{-1})_{kl} S_{lm} M_{mn},
\end{equation}
where
\begin{equation}
  (M) := 
  \begin{pmatrix}
    1 & 1 & 1 & 1 & 1 & 1 & 1 & 1 & 1 \\
    0 & 1 & 0 &-1 & 0 & 1 &-1 &-1 & 1 \\
    0 & 0 & 1 & 0 &-1 & 1 & 1 &-1 &-1 \\
    0 & 1 &-1 & 1 &-1 & 0 & 0 & 0 & 0 \\
    0 & 0 & 0 & 0 & 0 & 1 &-1 & 1 &-1 \\
   -4 &-1 &-1 &-1 &-1 & 2 & 2 & 2 & 2 \\
    0 &-2 & 0 & 2 & 0 & 1 &-1 &-1 & 1 \\
    0 & 0 &-2 & 0 & 2 & 1 & 1 &-1 &-1 \\
    4 &-2 &-2 &-2 &-2 & 1 & 1 & 1 & 1
  \end{pmatrix},
\end{equation}
is the matrix of moments~\cite{kruger2016lattice}. $(M^{-1})_{kl}$, is the inverse matrix, i.e.,
\begin{equation}
  \label{eq:moments_inverse}
  \sum_{l=0}^{Q-1} M_{kl} (M^{-1})_{lm} = \sum_{l=0}^{Q-1} (M^{-1})_{kl} M_{lm} = \delta_{kl},
\end{equation}
The matrix $S_{lm}$ contains the relaxation times, it is a diagonal matrix whose elements are given by
\begin{equation}
  \label{eq:relaxation_rates}
  \mathrm{diag} (S) := (1,\, 1,\, 1,\, \omega,\, \omega,\, 1,\, 1,\, 1,\, 1).
\end{equation}
where $\omega$ is specified by the dynamic viscosity~\cite{dHumieres2002mrt}, $\mu$,
\begin{equation}
  \label{eq:relax_time_visc}
  \omega = \left( \frac{\mu}{\rho c_s^2} + \frac{1}{2} \right)^{-1}.
\end{equation}

% section app:D2Q9 (end)

\section{The evolution of a distribution function}
\label{app:chapman-enskog}

Let us write the evolution of a distribution function, $f_q$, as
\begin{equation}
	f_q(\mathbfit{x} + \mathbfit{c}_q \Delta t, t + \Delta t) - f_q(\mathbfit{x}, t) = - \sum_{r=0}^{Q-1} \Lambda_{q r} [f_r - f_r^\text{eq}],
	\label{eq:lBe_potential_re}
\end{equation}
where we have included $\Delta t$ as a free parameter for the present derivation that can be later set to unity.
We now write the left-hand-side of Eq.~(\ref{eq:lBe_potential_re}) in a series expansion centered at $\Delta t = 0$, i.e.,
\begin{equation}
  \sum_{k=1}^{\infty}\frac{\Delta t^k}{k!} (\partial_t + \mathbfit{c}_q \cdot \nabla)^k f_q = - \sum_{r=0}^{Q-1} \Lambda_{q r} [f_r - f_r^\text{eq}].
  \label{eq:taylor-exp}
\end{equation}
This implies that, the equilibrium distribution function, $f_q^\text{eq}$, can be written in terms of $f_q$,
\begin{equation}
  f_q^\text{eq} = \left[ 1 + \mathbf{A} \right] f_q,
  \label{eq:equilibrium_dynamic}
\end{equation}
where we have defined the operator
\begin{equation}
  \label{eq:operator}
  \mathbf{A} := \sum_{r=0}^{Q-1}(\Lambda^{-1})_{qr} \sum_{k=1}^\infty \frac{1}{k!} \Delta t^k (\partial_t + \mathbfit{c}_r \cdot \nabla)^k,
\end{equation}
and $(\Lambda^{-1})_{qr}$ is the inverse of the collision matrix, which, from Eqs.~(\ref{eq:moments_inverse}) and (\ref{eq:relaxation_rates}), is given by,
\begin{equation}
  \label{eq:inverse_collision}
  (\Lambda^{-1})_{kn} = \sum_{l = 0}^{Q - 1} \sum_{m = 0}^{Q - 1} (M^{-1})_{k l} (S^{-1})_{l m} M_{m n},
\end{equation}
and, since $S$ is diagonal, the inverse is given by the reciprocal of its elements.

We can write an expression for $f_q$ in terms of $f_q^\text{eq}$ by inverting the relation in Eq.~(\ref{eq:equilibrium_dynamic}), i.e.,
\begin{equation}
  \label{eq:dynamics_equilibrium}
  f_q = \left[ 1 + \mathbf{A} \right]^{-1} f_q^\text{eq} = \left[ \sum_{m=0}^\infty (-1)^m \mathbf{A}^m \right] f_q^\text{eq}.
\end{equation}
In this way, the collision step can be expressed in terms of the equilibrium distribution,
\begin{equation}
  \label{eq:chapman-enskog_form}
  -\sum_{r=0}^{Q-1} \Lambda_{qr}[f_r - f_r^\text{eq}] = - \sum_{r=0}^{Q-1} \Lambda_{qr} \left[ \sum_{m=1}^\infty (-1)^m \mathbf{A}^m \right] f_r^\text{eq}.
\end{equation}
Substituting Eq.~(\ref{eq:operator}) and rearranging terms, Eq.~(\ref{eq:chapman-enskog_form}) becomes,
\begin{widetext}
  \begin{equation}
    -\sum_{r=0}^{Q-1} \Lambda_{qr} [f_r - f_r^\text{eq}] = \Delta t (\partial_t + \mathbfit{c}_q \cdot \nabla) f_q^\text{eq} + \frac{\Delta t^2 }{2} (\partial_t + \mathbfit{c}_q \cdot \nabla)^2 f_q^\text{eq} - \Delta t^2 (\partial_t + \mathbfit{c}_q \cdot \nabla) \sum_{r=0}^{Q-1} \left[ (\Lambda^{-1})_{qr} (\partial_t + \mathbfit{c}_r \cdot \nabla) f_r^\text{eq}  \right] \nonumber + O(\Delta t^3).
    \label{eq:red_chapman-enskog}
  \end{equation}
\end{widetext}
Finally, we now set $\Delta t = 1$, and for the case $\Lambda_{qr} = \delta_{qr}$ we obtain Eq.~(\ref{eq:Picard_lBe}).

% subsection app:chapman-enskog (end)

% section Appendix (end)

%%%END OF MAIN TEXT%%%

%%%REFERENCES%%%
% \bibliography{all}

\begin{thebibliography}{44}%
\makeatletter
\providecommand \@ifxundefined [1]{%
 \@ifx{#1\undefined}
}%
\providecommand \@ifnum [1]{%
 \ifnum #1\expandafter \@firstoftwo
 \else \expandafter \@secondoftwo
 \fi
}%
\providecommand \@ifx [1]{%
 \ifx #1\expandafter \@firstoftwo
 \else \expandafter \@secondoftwo
 \fi
}%
\providecommand \natexlab [1]{#1}%
\providecommand \enquote  [1]{``#1''}%
\providecommand \bibnamefont  [1]{#1}%
\providecommand \bibfnamefont [1]{#1}%
\providecommand \citenamefont [1]{#1}%
\providecommand \href@noop [0]{\@secondoftwo}%
\providecommand \href [0]{\begingroup \@sanitize@url \@href}%
\providecommand \@href[1]{\@@startlink{#1}\@@href}%
\providecommand \@@href[1]{\endgroup#1\@@endlink}%
\providecommand \@sanitize@url [0]{\catcode `\\12\catcode `\$12\catcode
  `\&12\catcode `\#12\catcode `\^12\catcode `\_12\catcode `\%12\relax}%
\providecommand \@@startlink[1]{}%
\providecommand \@@endlink[0]{}%
\providecommand \url  [0]{\begingroup\@sanitize@url \@url }%
\providecommand \@url [1]{\endgroup\@href {#1}{\urlprefix }}%
\providecommand \urlprefix  [0]{URL }%
\providecommand \Eprint [0]{\href }%
\providecommand \doibase [0]{http://dx.doi.org/}%
\providecommand \selectlanguage [0]{\@gobble}%
\providecommand \bibinfo  [0]{\@secondoftwo}%
\providecommand \bibfield  [0]{\@secondoftwo}%
\providecommand \translation [1]{[#1]}%
\providecommand \BibitemOpen [0]{}%
\providecommand \bibitemStop [0]{}%
\providecommand \bibitemNoStop [0]{.\EOS\space}%
\providecommand \EOS [0]{\spacefactor3000\relax}%
\providecommand \BibitemShut  [1]{\csname bibitem#1\endcsname}%
\let\auto@bib@innerbib\@empty
%</preamble>
\bibitem [{\citenamefont {Mugele}\ and\ \citenamefont
  {Baret}(2005)}]{mugele2005electrowetting}%
  \BibitemOpen
  \bibfield  {author} {\bibinfo {author} {\bibfnamefont {F.}~\bibnamefont
  {Mugele}}\ and\ \bibinfo {author} {\bibfnamefont {J.-C.}\ \bibnamefont
  {Baret}},\ }\href {http://stacks.iop.org/0953-8984/17/i=28/a=R01} {\bibfield
  {journal} {\bibinfo  {journal} {Journal of Physics: Condensed Matter}\
  }\textbf {\bibinfo {volume} {17}},\ \bibinfo {pages} {R705} (\bibinfo {year}
  {2005})}\BibitemShut {NoStop}%
\bibitem [{\citenamefont {Teh}\ \emph {et~al.}(2008)\citenamefont {Teh},
  \citenamefont {Lin}, \citenamefont {Hung},\ and\ \citenamefont
  {Lee}}]{teh2008droplet}%
  \BibitemOpen
  \bibfield  {author} {\bibinfo {author} {\bibfnamefont {S.-Y.}\ \bibnamefont
  {Teh}}, \bibinfo {author} {\bibfnamefont {R.}~\bibnamefont {Lin}}, \bibinfo
  {author} {\bibfnamefont {L.-H.}\ \bibnamefont {Hung}}, \ and\ \bibinfo
  {author} {\bibfnamefont {A.~P.}\ \bibnamefont {Lee}},\ }\href@noop {}
  {\bibfield  {journal} {\bibinfo  {journal} {Lab on a Chip}\ }\textbf
  {\bibinfo {volume} {8}},\ \bibinfo {pages} {198} (\bibinfo {year}
  {2008})}\BibitemShut {NoStop}%
\bibitem [{\citenamefont {Ren}\ \emph {et~al.}(2002)\citenamefont {Ren},
  \citenamefont {Fair}, \citenamefont {Pollack},\ and\ \citenamefont
  {Shaughnessy}}]{ren2002droplettransport}%
  \BibitemOpen
  \bibfield  {author} {\bibinfo {author} {\bibfnamefont {H.}~\bibnamefont
  {Ren}}, \bibinfo {author} {\bibfnamefont {R.~B.}\ \bibnamefont {Fair}},
  \bibinfo {author} {\bibfnamefont {M.~G.}\ \bibnamefont {Pollack}}, \ and\
  \bibinfo {author} {\bibfnamefont {E.~J.}\ \bibnamefont {Shaughnessy}},\
  }\href {\doibase https://doi.org/10.1016/S0925-4005(02)00223-X} {\bibfield
  {journal} {\bibinfo  {journal} {Sensors and Actuators B: Chemical}\ }\textbf
  {\bibinfo {volume} {87}},\ \bibinfo {pages} {201} (\bibinfo {year}
  {2002})}\BibitemShut {NoStop}%
\bibitem [{\citenamefont {Baret}\ \emph {et~al.}(2005)\citenamefont {Baret},
  \citenamefont {Decré}, \citenamefont {Herminghaus},\ and\ \citenamefont
  {Seemann}}]{baret2005electroactuation}%
  \BibitemOpen
  \bibfield  {author} {\bibinfo {author} {\bibfnamefont {J.-C.}\ \bibnamefont
  {Baret}}, \bibinfo {author} {\bibfnamefont {M.}~\bibnamefont {Decré}},
  \bibinfo {author} {\bibfnamefont {S.}~\bibnamefont {Herminghaus}}, \ and\
  \bibinfo {author} {\bibfnamefont {R.}~\bibnamefont {Seemann}},\ }\href
  {\doibase 10.1021/la052228b} {\bibfield  {journal} {\bibinfo  {journal}
  {Langmuir}\ }\textbf {\bibinfo {volume} {21}},\ \bibinfo {pages} {12218}
  (\bibinfo {year} {2005})},\ \bibinfo {note} {pMID: 16342995},\ \Eprint
  {http://arxiv.org/abs/https://doi.org/10.1021/la052228b}
  {https://doi.org/10.1021/la052228b} \BibitemShut {NoStop}%
\bibitem [{\citenamefont {Mugele}\ \emph {et~al.}(2006)\citenamefont {Mugele},
  \citenamefont {Baret},\ and\ \citenamefont
  {Steinhauser}}]{mugele2006electromixing}%
  \BibitemOpen
  \bibfield  {author} {\bibinfo {author} {\bibfnamefont {F.}~\bibnamefont
  {Mugele}}, \bibinfo {author} {\bibfnamefont {J.-C.}\ \bibnamefont {Baret}}, \
  and\ \bibinfo {author} {\bibfnamefont {D.}~\bibnamefont {Steinhauser}},\
  }\href {\doibase 10.1063/1.2204831} {\bibfield  {journal} {\bibinfo
  {journal} {Applied Physics Letters}\ }\textbf {\bibinfo {volume} {88}},\
  \bibinfo {pages} {204106} (\bibinfo {year} {2006})},\ \Eprint
  {http://arxiv.org/abs/https://doi.org/10.1063/1.2204831}
  {https://doi.org/10.1063/1.2204831} \BibitemShut {NoStop}%
\bibitem [{\citenamefont {Lu}\ \emph {et~al.}(2017)\citenamefont {Lu},
  \citenamefont {Sur}, \citenamefont {Pascente}, \citenamefont {Annapragada},
  \citenamefont {Ruchhoeft},\ and\ \citenamefont {Liu}}]{lu2017dynamics}%
  \BibitemOpen
  \bibfield  {author} {\bibinfo {author} {\bibfnamefont {Y.}~\bibnamefont
  {Lu}}, \bibinfo {author} {\bibfnamefont {A.}~\bibnamefont {Sur}}, \bibinfo
  {author} {\bibfnamefont {C.}~\bibnamefont {Pascente}}, \bibinfo {author}
  {\bibfnamefont {S.~R.}\ \bibnamefont {Annapragada}}, \bibinfo {author}
  {\bibfnamefont {P.}~\bibnamefont {Ruchhoeft}}, \ and\ \bibinfo {author}
  {\bibfnamefont {D.}~\bibnamefont {Liu}},\ }\href {\doibase
  https://doi.org/10.1016/j.ijheatmasstransfer.2016.10.040} {\bibfield
  {journal} {\bibinfo  {journal} {International Journal of Heat and Mass
  Transfer}\ }\textbf {\bibinfo {volume} {106}},\ \bibinfo {pages} {920 }
  (\bibinfo {year} {2017})}\BibitemShut {NoStop}%
\bibitem [{\citenamefont {Kudina}\ \emph {et~al.}(2016)\citenamefont {Kudina},
  \citenamefont {Eral},\ and\ \citenamefont {Mugele}}]{kudina2016emaldi}%
  \BibitemOpen
  \bibfield  {author} {\bibinfo {author} {\bibfnamefont {O.}~\bibnamefont
  {Kudina}}, \bibinfo {author} {\bibfnamefont {B.}~\bibnamefont {Eral}}, \ and\
  \bibinfo {author} {\bibfnamefont {F.}~\bibnamefont {Mugele}},\ }\href
  {\doibase 10.1021/acs.analchem.5b04283} {\bibfield  {journal} {\bibinfo
  {journal} {Analytical Chemistry}\ }\textbf {\bibinfo {volume} {88}},\
  \bibinfo {pages} {4669} (\bibinfo {year} {2016})},\ \bibinfo {note} {pMID:
  27026060},\ \Eprint
  {http://arxiv.org/abs/https://doi.org/10.1021/acs.analchem.5b04283}
  {https://doi.org/10.1021/acs.analchem.5b04283} \BibitemShut {NoStop}%
\bibitem [{\citenamefont {Schuhladen}\ \emph {et~al.}(2016)\citenamefont
  {Schuhladen}, \citenamefont {Banerjee}, \citenamefont {St{\"u}rmer},
  \citenamefont {M{\"u}ller}, \citenamefont {Wallrabe},\ and\ \citenamefont
  {Zappe}}]{schuhladen2016optofluidic}%
  \BibitemOpen
  \bibfield  {author} {\bibinfo {author} {\bibfnamefont {S.}~\bibnamefont
  {Schuhladen}}, \bibinfo {author} {\bibfnamefont {K.}~\bibnamefont
  {Banerjee}}, \bibinfo {author} {\bibfnamefont {M.}~\bibnamefont
  {St{\"u}rmer}}, \bibinfo {author} {\bibfnamefont {P.}~\bibnamefont
  {M{\"u}ller}}, \bibinfo {author} {\bibfnamefont {U.}~\bibnamefont
  {Wallrabe}}, \ and\ \bibinfo {author} {\bibfnamefont {H.}~\bibnamefont
  {Zappe}},\ }\href {http://dx.doi.org/10.1038/lsa.2016.5} {\bibfield
  {journal} {\bibinfo  {journal} {Light: Science \&Amp; Applications}\ }\textbf
  {\bibinfo {volume} {5}},\ \bibinfo {pages} {e16005 EP } (\bibinfo {year}
  {2016})},\ \bibinfo {note} {original Article}\BibitemShut {NoStop}%
\bibitem [{\citenamefont {Berge}\ and\ \citenamefont
  {Peseux}(2000)}]{berge2000electrolensing}%
  \BibitemOpen
  \bibfield  {author} {\bibinfo {author} {\bibfnamefont {B.}~\bibnamefont
  {Berge}}\ and\ \bibinfo {author} {\bibfnamefont {J.}~\bibnamefont {Peseux}},\
  }\href {\doibase 10.1007/s101890070029} {\bibfield  {journal} {\bibinfo
  {journal} {The European Physical Journal E}\ }\textbf {\bibinfo {volume}
  {3}},\ \bibinfo {pages} {159} (\bibinfo {year} {2000})}\BibitemShut {NoStop}%
\bibitem [{\citenamefont {Hayes}\ and\ \citenamefont
  {Feenstra}(2003)}]{hayes2003epaperspeed}%
  \BibitemOpen
  \bibfield  {author} {\bibinfo {author} {\bibfnamefont {R.~A.}\ \bibnamefont
  {Hayes}}\ and\ \bibinfo {author} {\bibfnamefont {B.~J.}\ \bibnamefont
  {Feenstra}},\ }\href {http://dx.doi.org/10.1038/nature01988} {\bibfield
  {journal} {\bibinfo  {journal} {Nature}\ }\textbf {\bibinfo {volume} {425}},\
  \bibinfo {pages} {383 EP } (\bibinfo {year} {2003})}\BibitemShut {NoStop}%
\bibitem [{\citenamefont {You}\ and\ \citenamefont
  {Steckl}(2010)}]{you2010epaper3color}%
  \BibitemOpen
  \bibfield  {author} {\bibinfo {author} {\bibfnamefont {H.}~\bibnamefont
  {You}}\ and\ \bibinfo {author} {\bibfnamefont {A.~J.}\ \bibnamefont
  {Steckl}},\ }\href {\doibase 10.1063/1.3464963} {\bibfield  {journal}
  {\bibinfo  {journal} {Applied Physics Letters}\ }\textbf {\bibinfo {volume}
  {97}},\ \bibinfo {pages} {023514} (\bibinfo {year} {2010})},\ \Eprint
  {http://arxiv.org/abs/https://doi.org/10.1063/1.3464963}
  {https://doi.org/10.1063/1.3464963} \BibitemShut {NoStop}%
\bibitem [{\citenamefont {Lomax}\ \emph {et~al.}(2016)\citenamefont {Lomax},
  \citenamefont {Kant}, \citenamefont {Williams}, \citenamefont {Patten},
  \citenamefont {Zou}, \citenamefont {Juel},\ and\ \citenamefont
  {Dryfe}}]{lomax2016ewoc}%
  \BibitemOpen
  \bibfield  {author} {\bibinfo {author} {\bibfnamefont {D.~J.}\ \bibnamefont
  {Lomax}}, \bibinfo {author} {\bibfnamefont {P.}~\bibnamefont {Kant}},
  \bibinfo {author} {\bibfnamefont {A.~T.}\ \bibnamefont {Williams}}, \bibinfo
  {author} {\bibfnamefont {H.~V.}\ \bibnamefont {Patten}}, \bibinfo {author}
  {\bibfnamefont {Y.}~\bibnamefont {Zou}}, \bibinfo {author} {\bibfnamefont
  {A.}~\bibnamefont {Juel}}, \ and\ \bibinfo {author} {\bibfnamefont
  {R.~A.~W.}\ \bibnamefont {Dryfe}},\ }\href {\doibase 10.1039/C6SM01565D}
  {\bibfield  {journal} {\bibinfo  {journal} {Soft Matter}\ }\textbf {\bibinfo
  {volume} {12}},\ \bibinfo {pages} {8798} (\bibinfo {year}
  {2016})}\BibitemShut {NoStop}%
\bibitem [{\citenamefont {Nelson}\ and\ \citenamefont
  {Kim}(2012)}]{nelson2012droplet}%
  \BibitemOpen
  \bibfield  {author} {\bibinfo {author} {\bibfnamefont {W.~C.}\ \bibnamefont
  {Nelson}}\ and\ \bibinfo {author} {\bibfnamefont {C.-J.}\ \bibnamefont
  {Kim}},\ }\href {\doibase 10.1163/156856111X599562} {\bibfield  {journal}
  {\bibinfo  {journal} {Journal of Adhesion Science and Technology}\ }\textbf
  {\bibinfo {volume} {26}},\ \bibinfo {pages} {1747} (\bibinfo {year}
  {2012})},\ \Eprint
  {http://arxiv.org/abs/https://www.tandfonline.com/doi/pdf/10.1163/156856111X599562}
  {https://www.tandfonline.com/doi/pdf/10.1163/156856111X599562} \BibitemShut
  {NoStop}%
\bibitem [{\citenamefont {Quilliet}\ and\ \citenamefont
  {Berge}(2002)}]{quilliet2002ewinterfacepot}%
  \BibitemOpen
  \bibfield  {author} {\bibinfo {author} {\bibfnamefont {C.}~\bibnamefont
  {Quilliet}}\ and\ \bibinfo {author} {\bibfnamefont {B.}~\bibnamefont
  {Berge}},\ }\href {http://stacks.iop.org/0295-5075/60/i=1/a=099} {\bibfield
  {journal} {\bibinfo  {journal} {EPL (Europhysics Letters)}\ }\textbf
  {\bibinfo {volume} {60}},\ \bibinfo {pages} {99} (\bibinfo {year}
  {2002})}\BibitemShut {NoStop}%
\bibitem [{\citenamefont {Kuo}\ \emph {et~al.}(2003)\citenamefont {Kuo},
  \citenamefont {Spicar-Mihalic}, \citenamefont {Rodriguez},\ and\
  \citenamefont {Chiu}}]{kuo2003ewmovement}%
  \BibitemOpen
  \bibfield  {author} {\bibinfo {author} {\bibfnamefont {J.~S.}\ \bibnamefont
  {Kuo}}, \bibinfo {author} {\bibfnamefont {P.}~\bibnamefont {Spicar-Mihalic}},
  \bibinfo {author} {\bibfnamefont {I.}~\bibnamefont {Rodriguez}}, \ and\
  \bibinfo {author} {\bibfnamefont {D.~T.}\ \bibnamefont {Chiu}},\ }\href
  {\doibase 10.1021/la020698p} {\bibfield  {journal} {\bibinfo  {journal}
  {Langmuir}\ }\textbf {\bibinfo {volume} {19}},\ \bibinfo {pages} {250}
  (\bibinfo {year} {2003})},\ \Eprint
  {http://arxiv.org/abs/https://doi.org/10.1021/la020698p}
  {https://doi.org/10.1021/la020698p} \BibitemShut {NoStop}%
\bibitem [{\citenamefont {Staicu}\ and\ \citenamefont
  {Mugele}(2006)}]{staicu2006entrapment}%
  \BibitemOpen
  \bibfield  {author} {\bibinfo {author} {\bibfnamefont {A.}~\bibnamefont
  {Staicu}}\ and\ \bibinfo {author} {\bibfnamefont {F.}~\bibnamefont
  {Mugele}},\ }\href {\doibase 10.1103/PhysRevLett.97.167801} {\bibfield
  {journal} {\bibinfo  {journal} {Phys. Rev. Lett.}\ }\textbf {\bibinfo
  {volume} {97}},\ \bibinfo {pages} {167801} (\bibinfo {year}
  {2006})}\BibitemShut {NoStop}%
\bibitem [{\citenamefont {Mugele}(2009)}]{mugele2009fundamental}%
  \BibitemOpen
  \bibfield  {author} {\bibinfo {author} {\bibfnamefont {F.}~\bibnamefont
  {Mugele}},\ }\href@noop {} {\bibfield  {journal} {\bibinfo  {journal} {Soft
  Matter}\ }\textbf {\bibinfo {volume} {5}},\ \bibinfo {pages} {3377} (\bibinfo
  {year} {2009})}\BibitemShut {NoStop}%
\bibitem [{\citenamefont {McHale}\ \emph {et~al.}(2013)\citenamefont {McHale},
  \citenamefont {Brown},\ and\ \citenamefont {Sampara}}]{mchale2013spreading}%
  \BibitemOpen
  \bibfield  {author} {\bibinfo {author} {\bibfnamefont {G.}~\bibnamefont
  {McHale}}, \bibinfo {author} {\bibfnamefont {C.~V.}\ \bibnamefont {Brown}}, \
  and\ \bibinfo {author} {\bibfnamefont {N.}~\bibnamefont {Sampara}},\ }\href
  {http://dx.doi.org/10.1038/ncomms2619} {\bibfield  {journal} {\bibinfo
  {journal} {Nature Communications}\ }\textbf {\bibinfo {volume} {4}} (\bibinfo
  {year} {2013})}\BibitemShut {NoStop}%
\bibitem [{\citenamefont {Buehrle}\ \emph {et~al.}(2003)\citenamefont
  {Buehrle}, \citenamefont {Herminghaus},\ and\ \citenamefont
  {Mugele}}]{buehrle2003profiles}%
  \BibitemOpen
  \bibfield  {author} {\bibinfo {author} {\bibfnamefont {J.}~\bibnamefont
  {Buehrle}}, \bibinfo {author} {\bibfnamefont {S.}~\bibnamefont
  {Herminghaus}}, \ and\ \bibinfo {author} {\bibfnamefont {F.}~\bibnamefont
  {Mugele}},\ }\href {\doibase 10.1103/PhysRevLett.91.086101} {\bibfield
  {journal} {\bibinfo  {journal} {Phys. Rev. Lett.}\ }\textbf {\bibinfo
  {volume} {91}},\ \bibinfo {pages} {086101} (\bibinfo {year}
  {2003})}\BibitemShut {NoStop}%
\bibitem [{\citenamefont {Quilliet}\ and\ \citenamefont
  {Berge}(2001)}]{quilliet2001cas}%
  \BibitemOpen
  \bibfield  {author} {\bibinfo {author} {\bibfnamefont {C.}~\bibnamefont
  {Quilliet}}\ and\ \bibinfo {author} {\bibfnamefont {B.}~\bibnamefont
  {Berge}},\ }\href {\doibase https://doi.org/10.1016/S1359-0294(00)00085-6}
  {\bibfield  {journal} {\bibinfo  {journal} {Current Opinion in Colloid \&Amp
  Interface Science}\ }\textbf {\bibinfo {volume} {6}},\ \bibinfo {pages} {34 }
  (\bibinfo {year} {2001})}\BibitemShut {NoStop}%
\bibitem [{\citenamefont {Kr{\"u}ger}\ \emph {et~al.}(2016)\citenamefont
  {Kr{\"u}ger}, \citenamefont {Kusumaatmaja}, \citenamefont {Kuzmin},
  \citenamefont {Shardt}, \citenamefont {Silva},\ and\ \citenamefont
  {Viggen}}]{kruger2016lattice}%
  \BibitemOpen
  \bibfield  {author} {\bibinfo {author} {\bibfnamefont {T.}~\bibnamefont
  {Kr{\"u}ger}}, \bibinfo {author} {\bibfnamefont {H.}~\bibnamefont
  {Kusumaatmaja}}, \bibinfo {author} {\bibfnamefont {A.}~\bibnamefont
  {Kuzmin}}, \bibinfo {author} {\bibfnamefont {O.}~\bibnamefont {Shardt}},
  \bibinfo {author} {\bibfnamefont {G.}~\bibnamefont {Silva}}, \ and\ \bibinfo
  {author} {\bibfnamefont {E.~M.}\ \bibnamefont {Viggen}},\ }\href@noop {}
  {\emph {\bibinfo {title} {The Lattice {B}oltzmann Method: Principles and
  Practice}}}\ (\bibinfo  {publisher} {Springer},\ \bibinfo {year}
  {2016})\BibitemShut {NoStop}%
\bibitem [{\citenamefont {Li}\ and\ \citenamefont {Fang}(2009)}]{li2009lbmew}%
  \BibitemOpen
  \bibfield  {author} {\bibinfo {author} {\bibfnamefont {H.}~\bibnamefont
  {Li}}\ and\ \bibinfo {author} {\bibfnamefont {H.}~\bibnamefont {Fang}},\
  }\href {\doibase 10.1140/epjst/e2009-01020-0} {\bibfield  {journal} {\bibinfo
   {journal} {The European Physical Journal Special Topics}\ }\textbf {\bibinfo
  {volume} {171}},\ \bibinfo {pages} {129} (\bibinfo {year}
  {2009})}\BibitemShut {NoStop}%
\bibitem [{\citenamefont {Clime}\ \emph {et~al.}(2010)\citenamefont {Clime},
  \citenamefont {Brassard},\ and\ \citenamefont
  {Veres}}]{clime2010numericalew}%
  \BibitemOpen
  \bibfield  {author} {\bibinfo {author} {\bibfnamefont {L.}~\bibnamefont
  {Clime}}, \bibinfo {author} {\bibfnamefont {D.}~\bibnamefont {Brassard}}, \
  and\ \bibinfo {author} {\bibfnamefont {T.}~\bibnamefont {Veres}},\ }\href
  {\doibase https://doi.org/10.1016/j.compfluid.2010.05.003} {\bibfield
  {journal} {\bibinfo  {journal} {Computers \&Amp Fluids}\ }\textbf {\bibinfo
  {volume} {39}},\ \bibinfo {pages} {1510 } (\bibinfo {year}
  {2010})}\BibitemShut {NoStop}%
\bibitem [{\citenamefont {Aminfar}\ and\ \citenamefont
  {Mohammadpourfard}(2009)}]{aminfar2009lbmew}%
  \BibitemOpen
  \bibfield  {author} {\bibinfo {author} {\bibfnamefont {H.}~\bibnamefont
  {Aminfar}}\ and\ \bibinfo {author} {\bibfnamefont {M.}~\bibnamefont
  {Mohammadpourfard}},\ }\href {\doibase
  https://doi.org/10.1016/j.cma.2009.08.021} {\bibfield  {journal} {\bibinfo
  {journal} {Computer Methods in Applied Mechanics and Engineering}\ }\textbf
  {\bibinfo {volume} {198}},\ \bibinfo {pages} {3852 } (\bibinfo {year}
  {2009})}\BibitemShut {NoStop}%
\bibitem [{\citenamefont {Lippmann}(1875)}]{lippmann1875relations}%
  \BibitemOpen
  \bibfield  {author} {\bibinfo {author} {\bibfnamefont {G.}~\bibnamefont
  {Lippmann}},\ }\emph {\bibinfo {title} {Relations entre les
  ph{\'e}nom{\`e}nes {\'e}lectriques et capillaires}},\ \href@noop {} {Ph.D.
  thesis},\ \bibinfo  {school} {Gauthier-Villars} (\bibinfo {year}
  {1875})\BibitemShut {NoStop}%
\bibitem [{\citenamefont {Landau}\ and\ \citenamefont
  {Lifshitz}(1980)}]{landau1980statistical}%
  \BibitemOpen
  \bibfield  {author} {\bibinfo {author} {\bibfnamefont {L.~D.}\ \bibnamefont
  {Landau}}\ and\ \bibinfo {author} {\bibfnamefont {E.~M.}\ \bibnamefont
  {Lifshitz}},\ }\href@noop {} {\emph {\bibinfo {title} {{Statistical Physics,
  Part 1}}}},\ \bibinfo {series} {Course of Theoretical Physics}, Vol.~\bibinfo
  {volume} {5}\ (\bibinfo  {publisher} {Butterworth-Heinemann},\ \bibinfo
  {address} {Oxford},\ \bibinfo {year} {1980})\BibitemShut {NoStop}%
\bibitem [{\citenamefont {Bray}(1994)}]{bray1994theory}%
  \BibitemOpen
  \bibfield  {author} {\bibinfo {author} {\bibfnamefont {A.~J.}\ \bibnamefont
  {Bray}},\ }\href {\doibase 10.1080/00018739400101505} {\bibfield  {journal}
  {\bibinfo  {journal} {Advances in Physics}\ }\textbf {\bibinfo {volume}
  {43}},\ \bibinfo {pages} {357} (\bibinfo {year} {1994})},\ \Eprint
  {http://arxiv.org/abs/http://dx.doi.org/10.1080/00018739400101505}
  {http://dx.doi.org/10.1080/00018739400101505} \BibitemShut {NoStop}%
\bibitem [{\citenamefont {Cahn}\ and\ \citenamefont
  {Hilliard}(1958)}]{cahn1958free-i}%
  \BibitemOpen
  \bibfield  {author} {\bibinfo {author} {\bibfnamefont {J.~W.}\ \bibnamefont
  {Cahn}}\ and\ \bibinfo {author} {\bibfnamefont {J.~E.}\ \bibnamefont
  {Hilliard}},\ }\href@noop {} {\bibfield  {journal} {\bibinfo  {journal} {The
  Journal of chemical physics}\ }\textbf {\bibinfo {volume} {28}},\ \bibinfo
  {pages} {258} (\bibinfo {year} {1958})}\BibitemShut {NoStop}%
\bibitem [{\citenamefont {Cahn}(1977)}]{cahn1977critical}%
  \BibitemOpen
  \bibfield  {author} {\bibinfo {author} {\bibfnamefont {J.~W.}\ \bibnamefont
  {Cahn}},\ }\href {\doibase http://dx.doi.org/10.1063/1.434402} {\bibfield
  {journal} {\bibinfo  {journal} {The Journal of Chemical Physics}\ }\textbf
  {\bibinfo {volume} {66}},\ \bibinfo {pages} {3667} (\bibinfo {year}
  {1977})}\BibitemShut {NoStop}%
\bibitem [{\citenamefont {Briant}\ \emph {et~al.}(2004)\citenamefont {Briant},
  \citenamefont {Wagner},\ and\ \citenamefont {Yeomans}}]{briant2004contact-i}%
  \BibitemOpen
  \bibfield  {author} {\bibinfo {author} {\bibfnamefont {A.~J.}\ \bibnamefont
  {Briant}}, \bibinfo {author} {\bibfnamefont {A.~J.}\ \bibnamefont {Wagner}},
  \ and\ \bibinfo {author} {\bibfnamefont {J.~M.}\ \bibnamefont {Yeomans}},\
  }\href {\doibase 10.1103/PhysRevE.69.031602} {\bibfield  {journal} {\bibinfo
  {journal} {Phys. Rev. E}\ }\textbf {\bibinfo {volume} {69}},\ \bibinfo
  {pages} {031602} (\bibinfo {year} {2004})}\BibitemShut {NoStop}%
\bibitem [{\citenamefont {Yang}\ \emph {et~al.}(1976)\citenamefont {Yang},
  \citenamefont {Fleming},\ and\ \citenamefont {Gibbs}}]{yang1976molecular}%
  \BibitemOpen
  \bibfield  {author} {\bibinfo {author} {\bibfnamefont {A.~J.~M.}\
  \bibnamefont {Yang}}, \bibinfo {author} {\bibfnamefont {P.~D.}\ \bibnamefont
  {Fleming}}, \ and\ \bibinfo {author} {\bibfnamefont {J.~H.}\ \bibnamefont
  {Gibbs}},\ }\href {\doibase http://dx.doi.org/10.1063/1.432687} {\bibfield
  {journal} {\bibinfo  {journal} {The Journal of Chemical Physics}\ }\textbf
  {\bibinfo {volume} {64}},\ \bibinfo {pages} {3732} (\bibinfo {year}
  {1976})}\BibitemShut {NoStop}%
\bibitem [{\citenamefont {Landau}\ \emph {et~al.}(2013)\citenamefont {Landau},
  \citenamefont {Bell}, \citenamefont {Kearsley}, \citenamefont {Pitaevskii},
  \citenamefont {Lifshitz},\ and\ \citenamefont
  {Sykes}}]{landau2013electrodynamics}%
  \BibitemOpen
  \bibfield  {author} {\bibinfo {author} {\bibfnamefont {L.~D.}\ \bibnamefont
  {Landau}}, \bibinfo {author} {\bibfnamefont {J.~S.}\ \bibnamefont {Bell}},
  \bibinfo {author} {\bibfnamefont {M.~J.}\ \bibnamefont {Kearsley}}, \bibinfo
  {author} {\bibfnamefont {L.~P.}\ \bibnamefont {Pitaevskii}}, \bibinfo
  {author} {\bibfnamefont {E.~M.}\ \bibnamefont {Lifshitz}}, \ and\ \bibinfo
  {author} {\bibfnamefont {J.~B.}\ \bibnamefont {Sykes}},\ }\href@noop {}
  {\emph {\bibinfo {title} {Electrodynamics of continuous media}}},\
  Vol.~\bibinfo {volume} {8}\ (\bibinfo  {publisher} {elsevier},\ \bibinfo
  {year} {2013})\BibitemShut {NoStop}%
\bibitem [{\citenamefont {Jackson}(1999)}]{jackson1999classical}%
  \BibitemOpen
  \bibfield  {author} {\bibinfo {author} {\bibfnamefont {J.~D.}\ \bibnamefont
  {Jackson}},\ }\href@noop {} {\emph {\bibinfo {title} {Classical
  electrodynamics}}}\ (\bibinfo  {publisher} {Wiley},\ \bibinfo {year}
  {1999})\BibitemShut {NoStop}%
\bibitem [{\citenamefont {Swift}\ \emph {et~al.}(1996)\citenamefont {Swift},
  \citenamefont {Orlandini}, \citenamefont {Osborn},\ and\ \citenamefont
  {Yeomans}}]{swift1996lattice}%
  \BibitemOpen
  \bibfield  {author} {\bibinfo {author} {\bibfnamefont {M.~R.}\ \bibnamefont
  {Swift}}, \bibinfo {author} {\bibfnamefont {E.}~\bibnamefont {Orlandini}},
  \bibinfo {author} {\bibfnamefont {W.~R.}\ \bibnamefont {Osborn}}, \ and\
  \bibinfo {author} {\bibfnamefont {J.~M.}\ \bibnamefont {Yeomans}},\
  }\href@noop {} {\bibfield  {journal} {\bibinfo  {journal} {Physical Review
  E}\ }\textbf {\bibinfo {volume} {54}},\ \bibinfo {pages} {5041} (\bibinfo
  {year} {1996})}\BibitemShut {NoStop}%
\bibitem [{\citenamefont {d{\textquoteright}Humi{\`e}res}\ \emph
  {et~al.}(2002)\citenamefont {d{\textquoteright}Humi{\`e}res}, \citenamefont
  {Ginzburg}, \citenamefont {Krafczyk}, \citenamefont {Lallemand},\ and\
  \citenamefont {Luo}}]{dHumieres2002mrt}%
  \BibitemOpen
  \bibfield  {author} {\bibinfo {author} {\bibfnamefont {D.}~\bibnamefont
  {d{\textquoteright}Humi{\`e}res}}, \bibinfo {author} {\bibfnamefont
  {I.}~\bibnamefont {Ginzburg}}, \bibinfo {author} {\bibfnamefont
  {M.}~\bibnamefont {Krafczyk}}, \bibinfo {author} {\bibfnamefont
  {P.}~\bibnamefont {Lallemand}}, \ and\ \bibinfo {author} {\bibfnamefont
  {L.-S.}\ \bibnamefont {Luo}},\ }\href {\doibase 10.1098/rsta.2001.0955}
  {\bibfield  {journal} {\bibinfo  {journal} {Philosophical Transactions of the
  Royal Society of London A: Mathematical, Physical and Engineering Sciences}\
  }\textbf {\bibinfo {volume} {360}},\ \bibinfo {pages} {437} (\bibinfo {year}
  {2002})},\ \Eprint
  {http://arxiv.org/abs/http://rsta.royalsocietypublishing.org/content/360/1792/437.full.pdf}
  {http://rsta.royalsocietypublishing.org/content/360/1792/437.full.pdf}
  \BibitemShut {NoStop}%
\bibitem [{\citenamefont {Desplat}\ \emph {et~al.}(2001)\citenamefont
  {Desplat}, \citenamefont {Pagonabarraga},\ and\ \citenamefont
  {Bladon}}]{desplat2001ludwig}%
  \BibitemOpen
  \bibfield  {author} {\bibinfo {author} {\bibfnamefont {J.~C.}\ \bibnamefont
  {Desplat}}, \bibinfo {author} {\bibfnamefont {I.}~\bibnamefont
  {Pagonabarraga}}, \ and\ \bibinfo {author} {\bibfnamefont {P.}~\bibnamefont
  {Bladon}},\ }\href {\doibase http://dx.doi.org/10.1016/S0010-4655(00)00205-8}
  {\bibfield  {journal} {\bibinfo  {journal} {Computer Physics Communications}\
  }\textbf {\bibinfo {volume} {134}},\ \bibinfo {pages} {273} (\bibinfo {year}
  {2001})}\BibitemShut {NoStop}%
\bibitem [{\citenamefont {Qian}\ and\ \citenamefont
  {Chen}(2000)}]{qian2000dissipative}%
  \BibitemOpen
  \bibfield  {author} {\bibinfo {author} {\bibfnamefont {Y.-H.}\ \bibnamefont
  {Qian}}\ and\ \bibinfo {author} {\bibfnamefont {S.-Y.}\ \bibnamefont
  {Chen}},\ }\href {\doibase 10.1103/PhysRevE.61.2712} {\bibfield  {journal}
  {\bibinfo  {journal} {Phys. Rev. E}\ }\textbf {\bibinfo {volume} {61}},\
  \bibinfo {pages} {2712} (\bibinfo {year} {2000})}\BibitemShut {NoStop}%
\bibitem [{\citenamefont {Ledesma-Aguilar}\ \emph {et~al.}(2014)\citenamefont
  {Ledesma-Aguilar}, \citenamefont {Vella},\ and\ \citenamefont
  {Yeomans}}]{ledesma2014lbmevaporation}%
  \BibitemOpen
  \bibfield  {author} {\bibinfo {author} {\bibfnamefont {R.}~\bibnamefont
  {Ledesma-Aguilar}}, \bibinfo {author} {\bibfnamefont {D.}~\bibnamefont
  {Vella}}, \ and\ \bibinfo {author} {\bibfnamefont {J.~M.}\ \bibnamefont
  {Yeomans}},\ }\href {\doibase 10.1039/C4SM01291G} {\bibfield  {journal}
  {\bibinfo  {journal} {Soft Matter}\ }\textbf {\bibinfo {volume} {10}},\
  \bibinfo {pages} {8267} (\bibinfo {year} {2014})}\BibitemShut {NoStop}%
\bibitem [{\citenamefont {Arfken}\ \emph {et~al.}(2013)\citenamefont {Arfken},
  \citenamefont {Weber},\ and\ \citenamefont
  {Harris}}]{arfken2013mathematical}%
  \BibitemOpen
  \bibfield  {author} {\bibinfo {author} {\bibfnamefont {G.~B.}\ \bibnamefont
  {Arfken}}, \bibinfo {author} {\bibfnamefont {H.~J.}\ \bibnamefont {Weber}}, \
  and\ \bibinfo {author} {\bibfnamefont {F.~E.}\ \bibnamefont {Harris}},\
  }\href@noop {} {\emph {\bibinfo {title} {Mathematical methods for Physicists,
  A Comprehensive guide, Seventh Edison}}}\ (\bibinfo  {publisher} {Elsevier,
  London},\ \bibinfo {year} {2013})\BibitemShut {NoStop}%
\bibitem [{\citenamefont {Yu}\ \emph {et~al.}(2003)\citenamefont {Yu},
  \citenamefont {Mei}, \citenamefont {Luo},\ and\ \citenamefont
  {S.}}]{yu2003viscous}%
  \BibitemOpen
  \bibfield  {author} {\bibinfo {author} {\bibfnamefont {D.}~\bibnamefont
  {Yu}}, \bibinfo {author} {\bibfnamefont {R.}~\bibnamefont {Mei}}, \bibinfo
  {author} {\bibfnamefont {L.-S.}\ \bibnamefont {Luo}}, \ and\ \bibinfo
  {author} {\bibfnamefont {W.}~\bibnamefont {S.}},\ }\href {\doibase
  http://dx.doi.org/10.1016/S0376-0421(03)00003-4} {\bibfield  {journal}
  {\bibinfo  {journal} {Progress in Aerospace Sciences}\ }\textbf {\bibinfo
  {volume} {39}},\ \bibinfo {pages} {329} (\bibinfo {year} {2003})}\BibitemShut
  {NoStop}%
\bibitem [{\citenamefont {Quinn}\ \emph {et~al.}(2005)\citenamefont {Quinn},
  \citenamefont {Sedev},\ and\ \citenamefont {Ralston}}]{quinn2005contact}%
  \BibitemOpen
  \bibfield  {author} {\bibinfo {author} {\bibfnamefont {A.}~\bibnamefont
  {Quinn}}, \bibinfo {author} {\bibfnamefont {R.}~\bibnamefont {Sedev}}, \ and\
  \bibinfo {author} {\bibfnamefont {J.}~\bibnamefont {Ralston}},\ }\href
  {\doibase 10.1021/jp040478f} {\bibfield  {journal} {\bibinfo  {journal} {The
  Journal of Physical Chemistry B}\ }\textbf {\bibinfo {volume} {109}},\
  \bibinfo {pages} {6268} (\bibinfo {year} {2005})},\ \bibinfo {note} {pMID:
  16851696},\ \Eprint {http://arxiv.org/abs/https://doi.org/10.1021/jp040478f}
  {https://doi.org/10.1021/jp040478f} \BibitemShut {NoStop}%
\bibitem [{\citenamefont {Peykov}\ \emph {et~al.}(2000)\citenamefont {Peykov},
  \citenamefont {Quinn},\ and\ \citenamefont
  {Ralston}}]{peykov2000electrowetting}%
  \BibitemOpen
  \bibfield  {author} {\bibinfo {author} {\bibfnamefont {V.}~\bibnamefont
  {Peykov}}, \bibinfo {author} {\bibfnamefont {A.}~\bibnamefont {Quinn}}, \
  and\ \bibinfo {author} {\bibfnamefont {J.}~\bibnamefont {Ralston}},\ }\href
  {\doibase 10.1007/s003960000333} {\bibfield  {journal} {\bibinfo  {journal}
  {Colloid and Polymer Science}\ }\textbf {\bibinfo {volume} {278}},\ \bibinfo
  {pages} {789} (\bibinfo {year} {2000})}\BibitemShut {NoStop}%
\bibitem [{\citenamefont {Oron}\ \emph {et~al.}(1997)\citenamefont {Oron},
  \citenamefont {Davis},\ and\ \citenamefont {Bankoff}}]{oron1997evolution}%
  \BibitemOpen
  \bibfield  {author} {\bibinfo {author} {\bibfnamefont {A.}~\bibnamefont
  {Oron}}, \bibinfo {author} {\bibfnamefont {S.~H.}\ \bibnamefont {Davis}}, \
  and\ \bibinfo {author} {\bibfnamefont {S.~G.}\ \bibnamefont {Bankoff}},\
  }\href {\doibase 10.1103/RevModPhys.69.931} {\bibfield  {journal} {\bibinfo
  {journal} {Rev. Mod. Phys.}\ }\textbf {\bibinfo {volume} {69}},\ \bibinfo
  {pages} {931} (\bibinfo {year} {1997})}\BibitemShut {NoStop}%
\bibitem [{\citenamefont {Andrew}\ \emph {et~al.}(2016)\citenamefont {Andrew},
  \citenamefont {Ledesma-Aguilar}, \citenamefont {Newton}, \citenamefont
  {Brown},\ and\ \citenamefont {McHale}}]{edwards2016notspreading}%
  \BibitemOpen
  \bibfield  {author} {\bibinfo {author} {\bibfnamefont {E.}~\bibnamefont
  {Andrew}}, \bibinfo {author} {\bibfnamefont {R.}~\bibnamefont
  {Ledesma-Aguilar}}, \bibinfo {author} {\bibfnamefont {M.}~\bibnamefont
  {Newton}}, \bibinfo {author} {\bibfnamefont {C.}~\bibnamefont {Brown}}, \
  and\ \bibinfo {author} {\bibfnamefont {G.}~\bibnamefont {McHale}},\ }\href
  {\doibase 10.1126/sciadv.1600183} {\bibfield  {journal} {\bibinfo  {journal}
  {Science Advances}\ }\textbf {\bibinfo {volume} {2}},\ \bibinfo {pages}
  {e1600183} (\bibinfo {year} {2016})}\BibitemShut {NoStop}%
\end{thebibliography}
% \input{manuscript_arXiv.bbl}
% \bibliographystyle{rsc}

%

\end{document}